\documentclass[preprint,showpacs,preprintnumbers,amsmath,amssymb,superscriptaddress]{revtex4}

\usepackage{graphicx}
\usepackage{dcolumn}
\usepackage{bm}

\usepackage{verbatim}

\usepackage{xcolor}
\definecolor{granata}{HTML}{831d1c}
\definecolor{kulblue}{HTML}{116E8A}

\DeclareMathOperator{\Tr}{Tr}
\DeclareMathOperator{\sech}{sech}
 \newcommand{\bfE}{\mathbf{E}}

\newcommand{\bfB}{\mathbf{B}}

\newcommand{\bfS}{\mathbf{S}}

\newcommand{\bfJ}{\mathbf{J}}

\newcommand{\bfv}{\mathbf{v}}

\newcommand{\bfV}{\mathbf{V}}

\newcommand{\bhx}{\hat{\mathbf{x}}}
\newcommand{\bhz}{\hat{\mathbf{z}}}

\newcommand{\alf}{{Alfv\'en }}

\begin{document}

\title{Local regimes of turbulence in 3D magnetic reconnection. }%
\author{G. Lapenta} 
\affiliation{Department of Mathematics, Center for mathematical Plasma Astrophysics, KU Leuven, University of Leuven, Belgium}
\author{F. Pucci} 
\affiliation{Department of Mathematics, Center for mathematical Plasma Astrophysics, KU Leuven, University of Leuven, Belgium}
\author{M.V. Goldman } 
\affiliation{Department of Physics, University of Colorado, Boulder, USA}
\author{D.L. Newman} 
\affiliation{Department of Physics, University of Colorado, Boulder, USA}

\date{\today}

\begin{abstract}
The process of magnetic reconnection when studied in Nature or when modeled in 3D simulations differs in one key way from the standard 2D paradigmatic cartoon: it is accompanied by much fluctuations in the electromagnetic fields and plasma properties. We developed a  diagnostics to study  the spectrum of fluctuations in the various regions around a reconnection site. We define the regions in terms of the local value of the flux function that determines the distance form the reconnection site, with positive values in the outflow and negative values in the inflow. We find that fluctuations belong to two very different regimes depending on the local plasma beta (defined as the ratio of plasma and magnetic pressure). The first regime develops in the reconnection outflows where beta is high and is characterized by a strong link between plasma and electromagnetic fluctuations leading to  momentum and energy  exchanges via anomalous viscosity and resistivity. But there is a second, low beta regime: it develops in the inflow and in the region around the separatrix surfaces, including the  reconnection electron diffusion region itself. It is remarkable that this low beta plasma,  where the magnetic pressure dominates, remain laminar even though the electromagnetic fields are turbulent. 
\end{abstract}

\maketitle

%
%

%


%
%
%
%

\section{Introduction}
Magnetic reconnection is a process believed to release large amounts of magnetic energy into kinetic energy \cite{zweibel2016perspectives} in system ranging from tokamaks (e.g. during disruptions) \cite{wesson1990sawtooth} to astrophysical flows \cite{sironi2014relativistic} to solar eruptions \cite{wyper2017universal} and geomagnetic storms \cite{birn2001geospace} to  experiments especially designed to study it
\cite{yamada2010magnetic}. Recently the magnetospheric multiscale (MMS) mission brought the field to an unprecedented level of resolution thanks to high cadence observations~\cite{burch2016electron}.
A key finding of the new mission is the confirmation that reconnection is associated with the development of strong electromagnetic fluctuations \cite{ergun2016magnetospheric,ergun2017drift}.

Previous observations from earlier missions \cite{retino2007situ, eastwood2009observations} and laboratory experiments~\cite{ji2004electromagnetic} had  already pointed to a link between reconnection and turbulence. One of the longest-known instabilities connected with reconnection is the lower hybrid drift instability (LHDI), long suspected to play a role in promoting reconnection \cite{huba1977lower} and observed in space \cite{vaivads2004cluster}  and in laboratory\cite{carter2001measurement}. The LHDI is driven by drifts caused by the presence of pressure (either density or temperature or both) gradients~\cite{huba1977lower}. As such the LHDI is  potentially present in the vicinity of reconnection, in regions where gradients are present and it has indeed been observed in reconnecting current layers~\citep{vaivads2004cluster,norgren2012lower}.

If reconnection can be caused by turbulence, reconnection is also known 
to produce turbulence \cite{karimabadi2013recent}. In 2D simulations, turbulence can be generated in the reconnection site itself via the tearing \cite{galeev1976tearing} and the plasmoid instability \cite{loureiro2007instability} and in the region around the  separatrices where electron holes form and slide rapidly \cite{cattell2005cluster,Divin2012,fujimoto2014wave, lapenta2014separatrices}. When the third dimension is added, the instabilities possible increase \cite{scholer2003onset,fujimoto2011reconnection}. 

In the reconnection outflows, the presence of gradients and currents allows the development of drift instabilities like the LHDI \cite{yin2008three,divin2015evolution,divin2015lower,price2017turbulence, le2017enhanced}. In the outflow region the strong density gradients have an unfavorable magnetic field curvature leading to  ballooning and interchange instabilities \cite{pritchett2011plasma,nakamura2016three,lapenta2011self} and flapping motions \cite{vapirev2013formation,sitnov2014magnetic}.   

The separatrix regions develop strong currents that can produce filamentation \cite{che2011current}, drift tearing \cite{Daughton:2011} and instabilities triggered in the separatrix density cavities \cite{markidis2012three} and velocity shears \cite{fermo2012secondary}. In the separatrices electron holes are driven by the streaming instabilities produced by strong electron beams~\cite{cattell2005cluster,lapenta2011bipolar,Goldman2014}. The electron holes can themselves become the source of whistler waves produced by {\v C}erenkov emission \citep{Goldman2014} that travel into the inflow region. Whistler waves can also be emitted by temperature anisotropy instabilities \citep{gary2006linear} in the outflow of reconnection \citep{fujimoto2008whistler}.

A practical consequence  of these fluctuations is the creation of anomalous momentum exchanges via viscosity and resistivity within the generalized Ohm's law, as recently reported \cite{price2017turbulence}.




In the present work, we consider the different regimes of fluctuation present in different regions around a reconnection site. To reach this goal we developed an investigation method  designed to provide statistical information on the fluctuation spectra in different regions. 
We identify two regimes of fluctuations. One in the outflow leads to a turbulent regime  where the fluctuations involve both fields and particles. In the inflow and separatrix region, instead, the fluctuations involve only the fields without affecting the particles significantly. 

The two regimes differ much in practical consequences. The outflow regime is capable of inducing a strong and turbulent energy exchange as well as strong anomalous momentum exchange dominated primarily by the electrostatic term in Ohm's law. The inflow regime, instead, does not lead to substantial fluctuations in the field-particle energy exchange nor significant  anomalous viscosity or resistivity. Turbulence remains limited to the electromagnetic fields. 


The work is organized as follows. Section \ref{methodology}  describes the simulation approach used and the diagnostics developed to analyze it; we focus in particular on a conditional statistical fluctuation analysis that identifies the properties in different regions. Section \ref{regimes} describes the main funding of the present work: the existence of two distinct regimes in the energy transfer due to the turbulent fluctuations. Section \ref{origin} discusses the origins and the causes of the two turbulent regimes observed. Section \ref{ohm} analyses the consequences of the fluctuation regimes on the momentum exchange via the generalized Ohm's law. Summary and Conclusions are drawn in Sect.~\ref{summary}.

\section{Methodology}
\label{methodology}
\begin{figure}[htbp]
\includegraphics[width=.9\columnwidth]{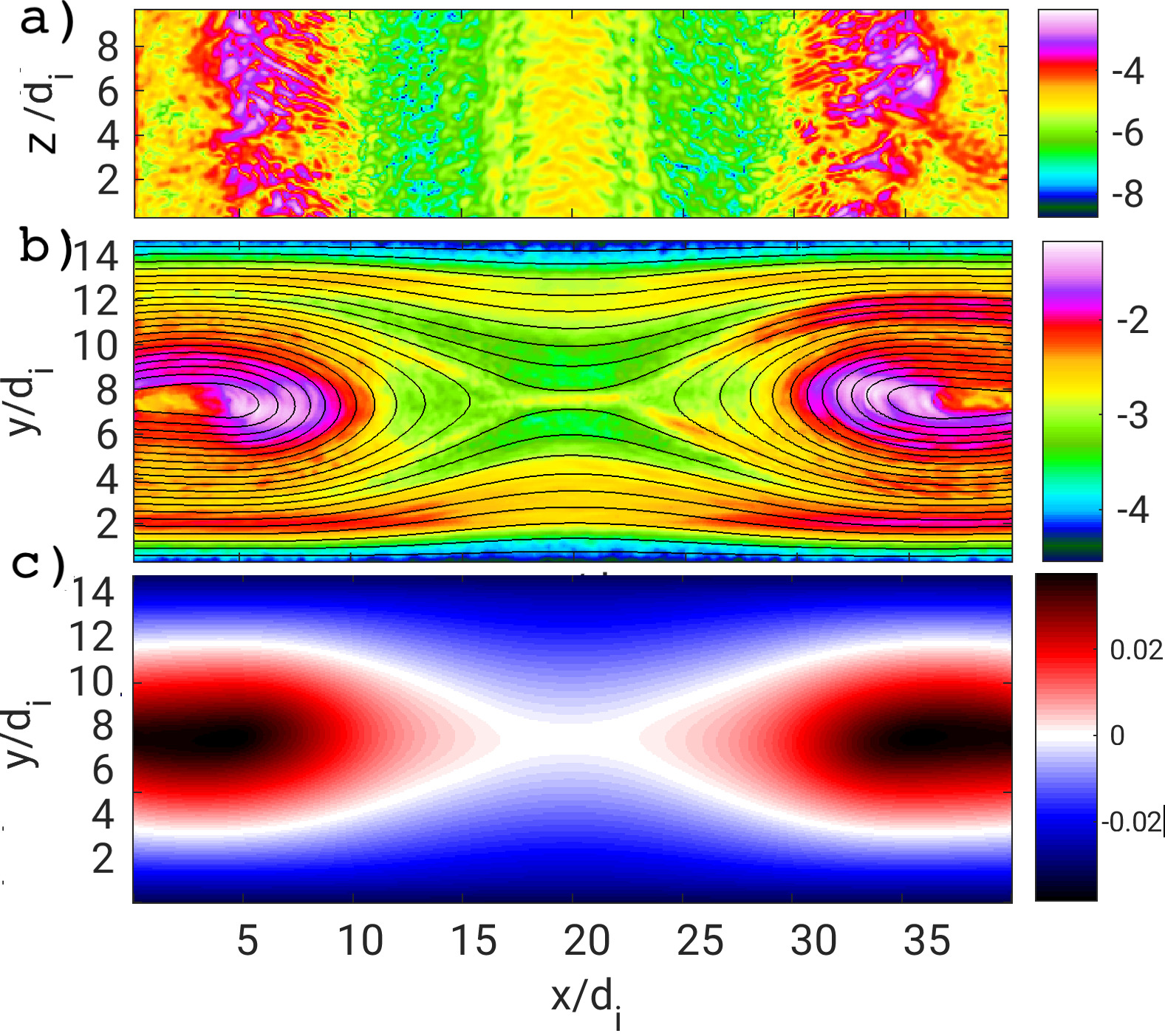}
\caption{Divergence of the Poynting flux at $\omega_{ct}=21.825$.  From top to bottom are reported: a) false color representation of $c\nabla \cdot \bfS/m_i\omega_{pi}^4$ in the mid-plane $y=L_y/2$; b)  the standard deviation of  the divergence of the Poynting flux, $ \langle (\nabla \cdot \bfS -\langle \nabla \cdot \bfS \rangle_z)^2 \rangle^{1/2} $, superimposed with the contour plot of the mean flux function $e\Phi/cm_i$;  c) 
flux function $e\Phi(x,y)/m_ic$ averaged over $z$. }
\label{figure_cuts}
\end{figure}
Our investigation uses the 3D fully kinetic particle in cell code iPic3D \cite{markidis2010multi}. The present simulation is similar to the one reported in \citep{lapenta2015secondary}, it uses a standard Harris equilibrium \cite{Harris1962}:
\begin{gather}
    \bfB(y)=B_0 \tanh(y/L) \bhx + B_g \bhz \\
    n(y)=n_0\sech^2(y/L)+n_b
\end{gather}
defined by the thickness $L/d_i=0.5$ and with the parameters $m_i/m_e=256$, $v_{the}/c=0.045$, $T_i/T_e=5$. A guide field $B_g/B_0=0.1$ and a background plasma of $n_b/n_0=0.1$ is imposed, where $B_0$ is the asymptotic in plane  field and $n_0$ is the peak Harris density. This choice of initial equilibrium corresponds to plasma beta (ratio of plasma pressure and magnetic pressure) peaking in the center of the sheet:
\begin{equation}
    \beta=\frac{8\pi n_0k(T_e+T_i)}{B^2_0}\left(\frac{\sech^2(y/L) +n_b/n_0}{\tanh^2(y/L)+B_g^2/B_0^2}\right)
\end{equation}
Using the parameters listed above, the plasma $\beta$ ranges from 11 in the center to 0.09 in the edge, covering 3 orders of magnitude.
These conditions are common in space plasmas, for example in the Earth magnetotail where both density and temperature peak in the center of the current layer~\citep{runov2006local}, but also in the solar wind~\citep{eriksson2014signatures}.

The coordinates are chosen with  the initial  Harris magnetic field  along $x$ with size  $L_x=40d_i$, the initial gradients  along $y$ with $L_y=15d_i$. The third dimension, where the initial current and guide field are directed,  is initially invariant with $L_z=10d_i$. Open boundaries are imposed in $x$ and $y$ and periodicity is assumed along $z$. 

 Unlike the simulation reported in reported in \citep{lapenta2015secondary, pucci2017properties} where there was a relative drift between the particles in the Harris equilibrium and in the background, we choose to set the simulation initially  in the ion rest frame so that all the current is initially carried by the electrons:
 \begin{gather}
    f_e=n_0 C_e \exp(-(\bfv-\bfv_{e0})^2/2v^2_{the}) +   n_b \exp(-v^2/2v^2_{the}) \\
     f_i=(n_0 +n_b) C_i \exp(-v^2/2v^2_{thi}) 
 \end{gather}
 where $C_s=(m/2\pi kT_s)^{3/2}$ is the normalization coefficient of the Maxwellian distribution and $\bfv_{e0}=-2ck(T_e+T_i)/eB_0L$ is the drift speed of the electrons required to support the Harris current when the simulation frame corresponds to the ion frame.
 In this frame the background plasma and the Harris ions are not drifting and the overall system has no velocity shear (since the electron mass is much smaller and the center of mass speed is essentially the speed of the ions that is initially zero). This choice avoids the presence of shear-driven modes that tend to kink the current sheet, an effect that would have complicated the interpretation of the fluctuations \citep{daughton02,lapenta2003unexpected,karimabadi2003ion}. The four plasma species (ions and electrons in the Harris and background) are each described by 125 particles per cell, with non uniform weight in the case of the Harris species. 
The grid  has 512x192x128 cells, resolving well the electron skin depth, in the reconnecting background plasma $\Delta  x \approx 0.4 d_{e,b}$ (the initial Harris plasma is quickly swept away by reconnection). The time step also resolves also well the electron cyclotron frequency, even in the strongest field, $\omega_{ce,Bmax}\Delta t \approx 0.3$ (and even better in more average fields). 

Figure \ref{figure_cuts} shows an overview of the simulation. The divergence of the Poynting flux is shown in the $y=L_y/2$  plane (panel a) to highlight  the fluctuations in the outflow. The standard deviation of the same flux  computed using the values along $z$, for each point in the x-y plane, shows several regions of  fluctuations also at the separatrices and the inflow, although at lower intensity than the outflow (panel b). We can distinguish three regions: 1) the inflow region above and below the central reconnection region (blue in Fig. \ref{figure_cuts}-c), 2) the central reconnection region and the  band around the separatrices (white in Fig. \ref{figure_cuts}-c), 3) the outflow region where the downstream pileup front form (red in Fig. \ref{figure_cuts}-c).  




In a previous study \cite{pucci2017properties} based on a similar 3D reconnection simulation (with different parameters), we investigated all these fluctuations collectively determining the spectrum to have a power law distribution compatible with a turbulent cascade and with the same index observed in space \cite{eastwood2009observations}. However, the fluctuations are far from homogeneous and visual inspection clearly points to different types of fluctuations in different parts of the domain. To analyze the fluctuations in each region, we have developed a \textit{conditional fluctuation analysis (CFA)}. The idea is to use the same statistical tools used in demographics. A statistician might pose the question of what is the income distribution among the people living in different districts of a city. Similarly we study the spectrum of fluctuations in different regions around the reconnection region. 

The main question is how do we identify the regions? The flux function is a natural choice, considering that the simulation is initially invariant in $z$ and throughout the evolution this invariance remains valid on average, except for the fluctuations. We define in each plane at constant $z$ a flux function $\Phi(x,y)$, defined as the scalar function whose contours are everywhere parallel to the magnetic field on that plane (i.e. the function is defined by the generating equation $\nabla_{xy} \Phi \cdot \bfB=0)$. Note that this function is not the out of plane component of the vector potential: this property will be true only in a 2D domain. In our 3D case, the flux function defined above is just a useful function to define the intersection of the magnetic field surfaces with a plane at given $z$. 

The flux function is defined minus a constant on each plane: we define it to be zero  in each plane  at the intersection where the separatrix surfaces meet $\Phi(x_{x},y_{x})=0$ (where $x_{x},y_{x}$ identifies the  so-called x-point in each plane, defined as the saddle point of $\Phi(x,y)$). Figure \ref{figure_cuts}-c reports an example of such flux function for the central plane ($z=Lz/2$). It obviously resembles the out of plane component of the vector potential in 2D domains despite the warning above. Similar plots are obtained in each plane along $z$. 

To define the  CFA, we subdivide the range of  $\Phi$ into 100 3D regions between two surfaces at two consecutive values of $\Phi$. The regions are 3D because $\Phi$ is defined in every $z$ plane. The region  around $\Phi=0$, by construction, is centered on the reconnection site. We then measure in each of these regions how the fluctuations for a given quantity are statistically distributed and report the histogram of the fluctuations with respect to the  mean in each region using a color scale. 

 

\begin{figure}[htbp]
\centering
\includegraphics[width=.9\columnwidth]{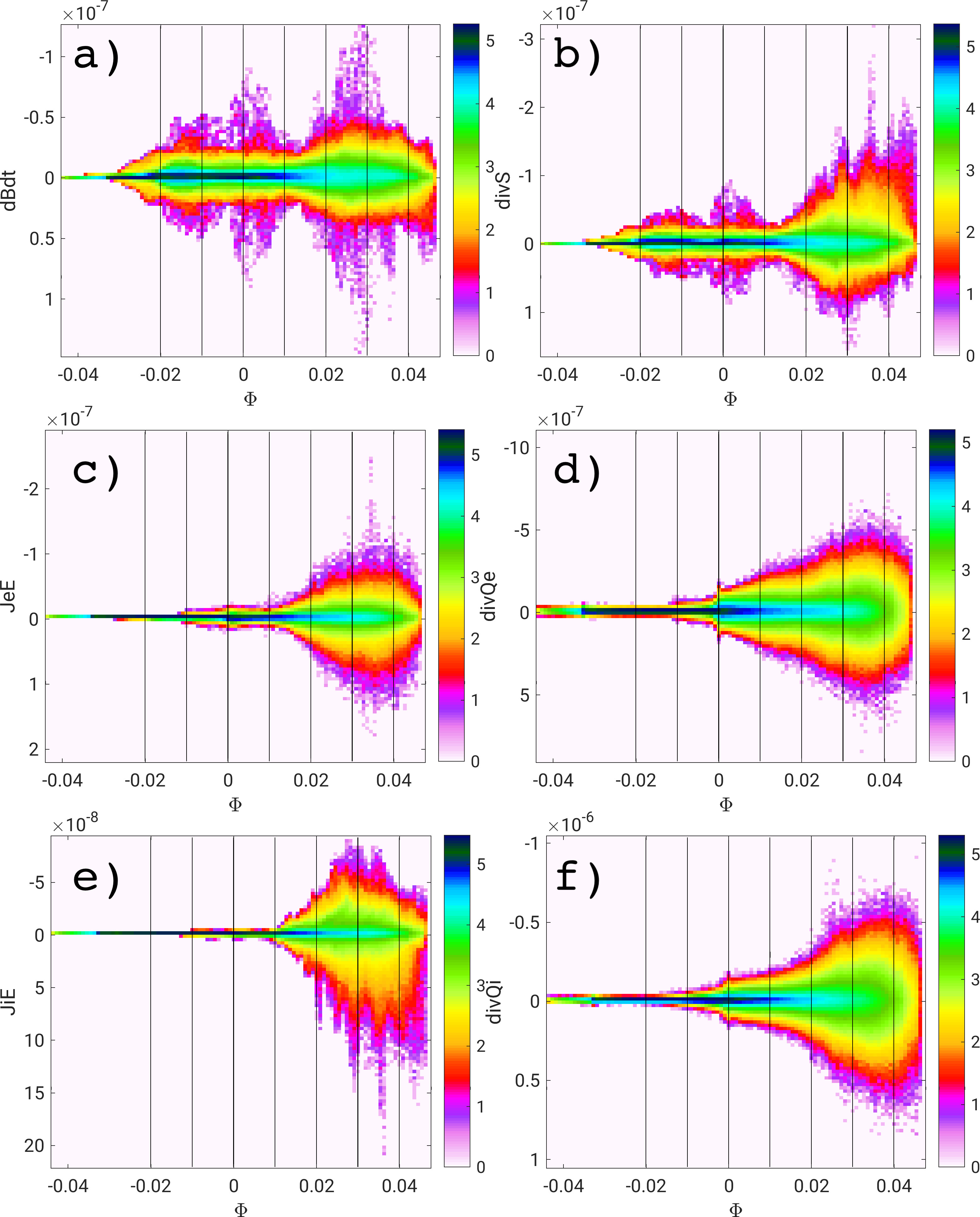}
\caption{Results of the CFA analysis displaying the occurrence count in 100 fluctuation ranges for 100 flux function domains. The quantities measured from top to bottom and left to right are: a) magnetic energy ($W_B=e^2B^2/2\mu_0 m_i^2 \omega_{pi}^2$) change $\omega_{pi}^{-1}\partial W_B/\partial t$, b) divergence of the Poynting flux $c\nabla \cdot \bfS/m_i \omega_{pi}^4$, c) $c\bfJ_e \cdot \bfE /m_i \omega_{pi}^4$, d) divergence of the electron energy flux, e) $c\bfJ_i \cdot \bfE /m_i \omega_{pi}^4$, f) divergence of the ion energy flux.}
\label{figure_cfa}
\end{figure}

\section{Two regimes of turbulence in proximity of a reconnection site}
\label{regimes}

The result of the analysis  described above is shown in Fig. \ref{figure_cfa}. What can be observed is the fluctuation range, defined with a sign because fluctuations can be positive or negative relative to the mean. In each panel we can observe the fluctuation spectrum in the region where the separatrices meet (around  $\Phi=0$), in the inflow ($\Phi<0$) and in the outflow ($\Phi>0$). The precise location corresponding to a given value of $\Phi$ can be obtained form Fig.~\ref{figure_cuts}-c. 

From the top, we report first the electromagnetic energy exchange between Poynting flux and magnetic energy (the electric field energy is negligible in our simulation where the speeds remain much below the speed of light). Next we report the electron energy exchange with the electromagnetic field and the electron particle energy flux (inclusive of all energy fluxes: bulk, enthalpy and heat flux). Finally the ions are shown. 

The most striking feature is that in the inflow ($\Phi<0$) and reconnection region proper ($\Phi\approx 0$) almost only the electromagnetic energy channels fluctuate. In essence the fluctuations only transfer energy between magnetic and Poynting flux, without affecting ions and electrons. In the downstream region ($\Phi>0$), instead, all energy exchanges fluctuate together. The  energy exchange rate with the electric field, $\bfJ_s \cdot \bfE$, and the particle energy fluxes fluctuate only in the outflow. 


The CFA clearly identifies two different regimes for the fluctuations. The first is in the region of inflow  ($\Phi<0$) where only the electromagnetic fields fluctuate but the particle energy exchange with the fields ($\bfJ_{s} \cdot \bfE$) does not respond to the fluctuations. The second region is that of the outflow ($\Phi>0$) where the species start to respond to the fluctuations. 

\begin{figure}[htbp]
\includegraphics[width=\columnwidth]{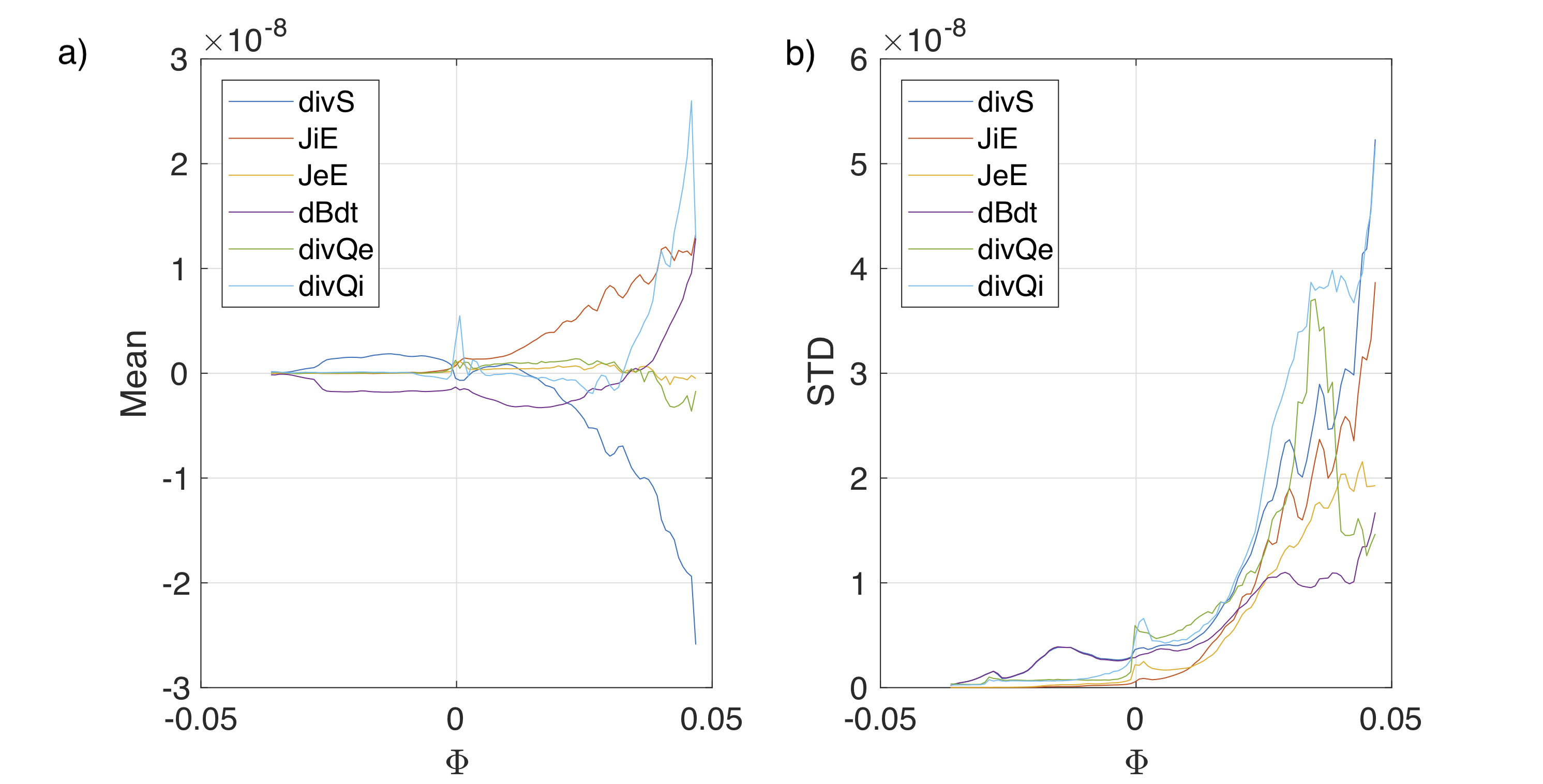}
\caption{Mean (panel a)  and standard deviation (STD, panel b) of the terms of the electromagnetic energy equation: divergence of the Poynting flux, energy exchange of the ions and electrons with the electric field ($c\bfJ_s \cdot \bfE /m_i \omega_{pi}^4$), divergence of the energy fluxes by species and temporal variation of the magnetic energy ($dW_B/\omega_{pi}dt$, shortened as  dB/dt  in the legend).}
\label{figure_std_mean}
\end{figure}

The fluctuations can be characterized by their mean and standard deviation (STD), reported in Fig.~\ref{figure_std_mean}. The region of the inflow is characterized by a strong mean and standard deviation for the magnetic energy and divergence of the Poynting flux but almost no response at all from the particle, neither as a mean nor as a STD. Right at the region $\Phi=0$, corresponding to the separatrices the divergence of the ion energy flux has a significant spike. This effect is caused by the Hall electric field present at the separatrices that accelerates the ions~\citep{aunai2011energy}.  Conversely, the outflow region involves a large mean and STD for all quantities, in fact the ion energy exchange term becomes dominant. 

\begin{figure}[htbp]
\includegraphics[width=\columnwidth]{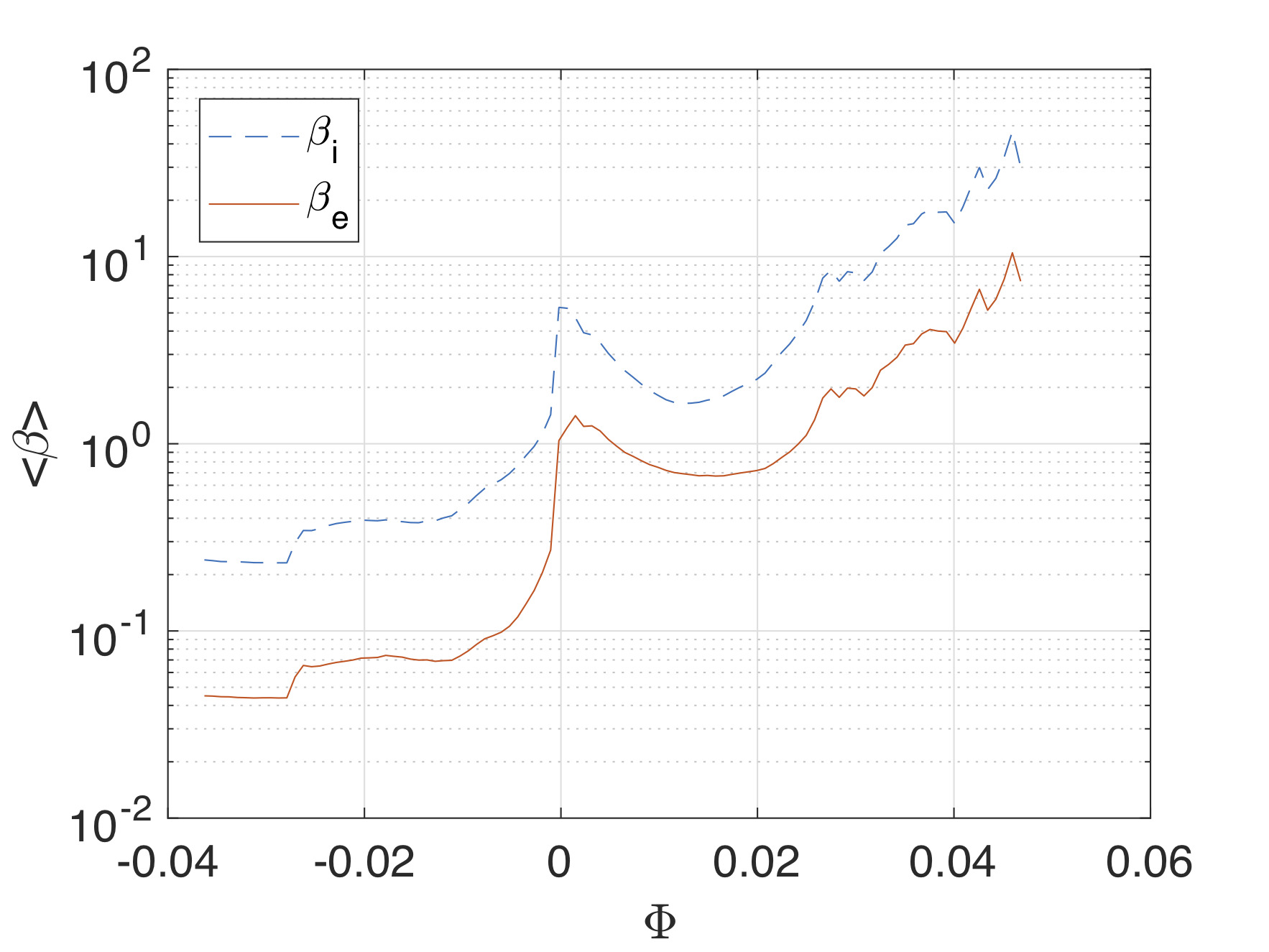}
\caption{Mean ion and electron plasma beta as a function of the flux function $\Phi$ defining the different  regions around a reconnection site.}
\label{figure_beta}
\end{figure}

These two regimes of fluctuation are linked with two different physical conditions. Figure \ref{figure_beta} shows the mean plasma beta for electrons and ions in the different regions defined above via $\Phi$. The inflow region has a low beta, lower than one. The outflow region has plasma local beta exceeding, even largely exceeding, unity. Beta for a species  is defined as:
\begin{equation}
\beta_s=\frac{\Tr  P_s}{B^2/(2\mu_0)}
\end{equation}
or the ratio of the species  pressure defined as the trace of the pressure tensor and the magnetic field pressure. In the inflow region the magnetic field pressure dominates largely and the turbulence remains mostly limited to the electromagnetic field. In the outflow, the plasma pressure becomes comparable or even exceeds the magnetic pressure and turbulence engulfs both particles and electromagnetic fields.

%
%


\section{Origin of the fluctuations}
\label{origin}
The system is initially laminar and reconnection is initiated in the same way in all planes $z$. In itself reconnection would progress in a laminar way without any variation of the quantities along $z$. But as noted above, the conditions generated by reconnection produce secondary instabilities that lead to fluctuations and eventually in the non linear phase of these instabilities to a turbulent spectrum. The non-linearity of these process complicates the precise determination of the linear instabilities first developing. We have in fact a system where the state is constantly altered by the reconnection process and the onset of secondary instabilities cannot be identified and compared with a precise linear theory analysis.

Nevertheless the fluctuation spectrum retains a memory of its cause that can be identified by measuring the local temporal evolution of the fields. We distribute uniformly within the simulation probes disposed in an array of 3 probes per dimensions in each processor of the simulation, for a total of 48x36x24 virtual probes. On these probes we save the electromagnetic fields and the species moments at all time steps. This information is complementary to the grid data. If the grid data gives the information at all points but for practical disk space limitation can be saved only infrequently (once every 1000 time steps), the probe data is limited in spatial resolution but it has full temporal resolution.  The virtual probes represent a simulation analogy to a multi-spacecraft mission, except we can afford a virtual mission of 41472 spacecraft. 

\begin{figure}[htbp]
\begin{tabular}{c}
a) \\
\includegraphics[width=.6\columnwidth]{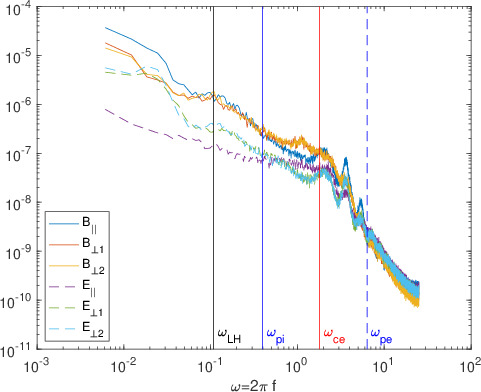}\\
 b)\\\includegraphics[width=.6\columnwidth]{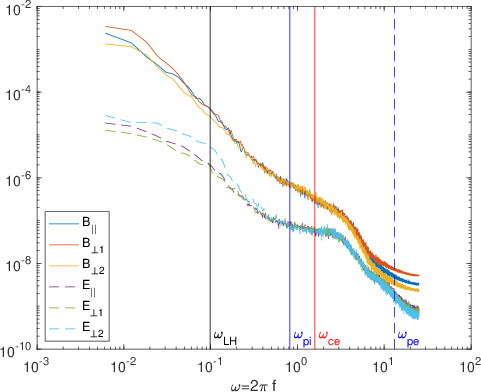} 
\end{tabular}
\caption{Fourier spectrum of the components of the electric ($e\bfE/cm_i\omega_{pi}$) and magnetic ($e\bfB/m_i\omega_{pi}$) field parallel and perpendicular to the mean magnetic field. The result is the ensemble average of 24 probes equally spaced along $z$ and positioned at $x/d_i=20.43$ and $y/d_i=10.20$ in the inflow (panel a) and at  $x/d_i=10.43$ and $y/d_i=7.70$ in the outflow (panel b). The data has been detrended by removing the linear interpolant between initial and final value.  The frequency is normalized to the ion plasma frequency computed with the initial Harris density. Some selected typical  frequencies are indicated by vertical lines and labeled.}
\label{figure_F_spectra}
\end{figure}

\begin{figure}[htbp]
\includegraphics[width=\columnwidth]{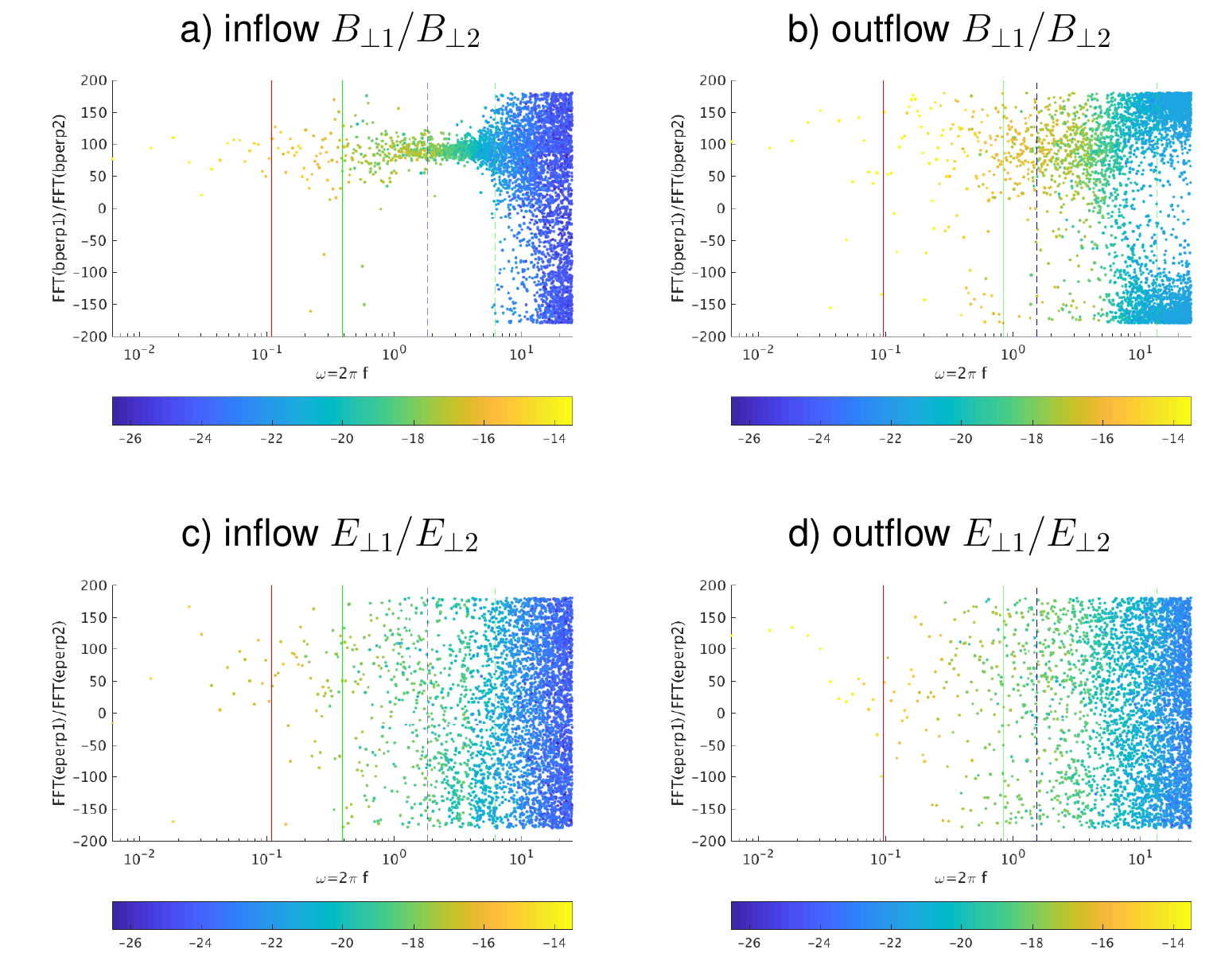}
\caption{Relative phase of the Fourier spectrum of the two perpendicular components of the magnetic field fluctuations, $\widetilde{B}_{\perp 1}/\widetilde{B}_{\perp 2}$ (top, panels a, b) and of the electric field fluctuations, $\widetilde{E}_{\perp 1}/\widetilde{E}_{\perp 2}$ (bottom, panels c,d). The inflow probes are shown on the left (panels a, c) and the outflow on the right (panels b, d).  The same locations and detrending are used as in Fig. \ref{figure_F_spectra}. Each frequency is reported with a dot in different color identified by the absolute value of the Fourier transformation  $\widetilde{B}_{\perp 1}$.  }
\label{figure_F_angle}
\end{figure}

To analyze the origin of the fluctuations in the inflow and outflow we consider one array of 24 virtual probes along $z$ at a given $(x,y)$ location. Figure~\ref{figure_F_spectra} shows two of these arrays, one in the inflow and one in the outflow. For each array of probes the average value of the local ion and electron plasma frequency of the lower hybrid frequency and electron cyclotron frequency are reported as vertical lines. 

The fluctuations of the electric and magnetic fields is computed relative to the direction of the average magnetic field measured by each probe, i.e. the fluctuations are projected with respect to the local time-averaged magnetic field and then averaged over the 24 probes. The spectra in the inflow and the outflow show remarkable differences. 

In the inflow, we observe a clear peak in the electron cyclotron frequency, $\omega_{ce}$. This peak is present primarily in the perturbations of the  perpendicular electric field  and in the parallel magnetic field, a direct indication that this wave is electromagnetic and propagates perpendicular to the direction of the average magnetic field. The two perpendicular components of the perturbations of $\bfB$ and $\bfE$ are nearly identical to each other, an indication of gyrotropy. Gyrotropy is broken only at very large frequencies where the electron gyromotion is not fast enough to re-equilibrate the two perpendicular components.    

A similar peak in the cyclotron frequency was reported in previous studies \citep{pucci2018generation}.  The frequencies reported are not precisely defined because the data provided is from 24 different probes and there is spatial and temporal variation. The value reported in  Fig.~\ref{figure_F_spectra} is an average. Nevertheless the correspondence of the peak with the cyclotron frequency is remarkable. Several higher harmonics are also evident. The size of the time step and time interval considered allows us to follow the entire range.

Perpendicular waves at the electron cyclotron frequency can be of different nature \cite{swanson}. The extraordinary mode is electromagnetic but centers at the electron plasma frequency and at the upper hybrid frequency, that is above what we observe here. Bernstein modes are concentrated at the scale observed but their nature is primarily electrostatic. We observe larger magnetic than electric perturbation in our units where the speed of light is unitary, this means the waves are electromagnetic, unlikely then to be Bernstein modes. However, recent MMS observations have reported peaks in the electron cyclotron range similar to those reported in Fig.~\ref{figure_F_spectra} and the authors suggest a Bernstein nature for these modes \citep{goodrich2018mms,wenya-mms}.

A different mechanism to generate multiple discrete peaks in the electron cyclotron to upper hybrid frequency range has been recently suggested \citep{dokgo-mms}. Based on MMS observations of waves in the upper hybrid \citep{burch2019high,dokgo2019high} and in the cyclotron range~\citep{goodrich2018mms,wenya-mms}, a model can be constructed based on the presence of a thermal plasma and a higher energy beam population spread over a range in pitch angles. The beam-plasma interaction leads to the generation of both type of waves of observed by MMS: upper hybrid or electron cyclotron. Multiple electron cyclotron peaks, similar to those observed in our study, are predicted by this theory when the beam density is sufficiently strong compared with that of the thermal plasma \citep{dokgo-mms}.  To resolve the issue a $k-\omega$ spectrum would be needed but we do  not have the resources to store all the time steps at all locations:  we only save the data at the virtual probe locations, precluding the calculation of the $k-\omega$ spectrum.

Just below the cyclotron peak, the perpendicular magnetic field perturbations become dominant over the parallel magnetic perturbation (and, conversely, for the electric field the parallel perturbation dominates over the perpendicular), indicating parallel propagation: this is a clear indication of whistler waves in the range below the electron cyclotron frequency, as noted in  \citet{pucci2018generation}.

At even lower frequencies, a peak at the lower hybrid frequency is present in the fluctuations of $\bfB$ and in the perpendicular perturbations of $\bfE$.  The lower hybrid drift instability (LHDI) is characteristically present in the interface region between the center of the current sheet and the lobes, even in absence of reconnection~\citep{huba1980lower}. The LHDI is a very common signature  observed in current layers in space  \citep{vaivads2004cluster,norgren2012lower} and in laboratory~\citep{carter2001measurement,ji2004electromagnetic} and covers a wide range of frequencies both linearly~\citep{daughton2003electromagnetic} and non-linearly~\citep{innocenti2016study}.

In the outflow, the electron cyclotron peak and its higher harmonics are completely absent. Now the spectrum is virtually identical for all three components of each field and no significant peak is present. Rather, the spectrum present the typical power law decay of turbulence. Note also the scale: the spectrum is much stronger in the outflow: the level of fluctuation is much higher and turbulence develops more fully than in the inflow. Previous studies \citep{vapirev2015initial,lapenta2015secondary} have shown that the turbulence in the outflow is generated by an instability in the lower hybrid range. A detailed comparison with liner theory shows that most of the growth rate observed can be explained by the lower hybrid range caused by the density gradients in the outflow but additional effects due to field line curvature and velocity shear play a role as well~\citep{divin2015evolution}. This instability has then been confirmed by satellite observations~\citep{divin2015lower}. Regardless of the origin, by the time reported, turbulence has developed fully, loosing memory of the scales that have generated it~\cite{innocenti2016study}.

\begin{figure}[htbp]
\includegraphics[width=.7\columnwidth]{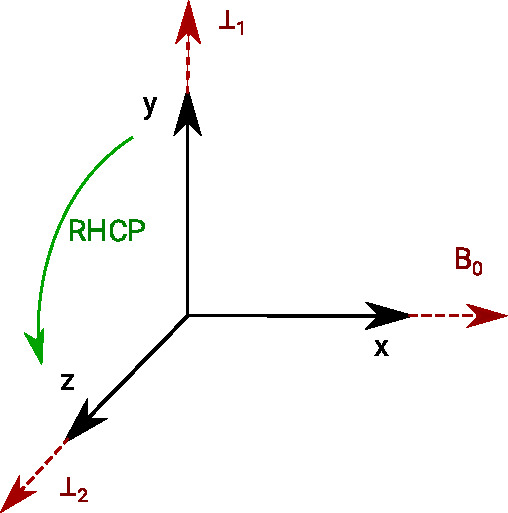}
\caption{Coordinate system and evolution of a right handed circularly polarized  (RHCP) wave.  The two perpendicular direction are indicated. The unit vector  $\hat{\mathbf{n}}_{\perp 1}$ is the projection of the perpendicular direction on the $xy$ plane: $\hat{\mathbf{n}}_{\perp 1} = (b_y, -b_x, 0)$, where the  magnetic field used as reference is averaged over all 24 probes and time: $\mathbf{b}=\langle \bfB \rangle / \langle B \rangle$.  The unit vector  $\hat{\mathbf{n}}_{\perp 2}$ is normal to both $\mathbf{b}$ and $\hat{\mathbf{n}}_{\perp 1}$: $\hat{\mathbf{n}}_{\perp 2}= \mathbf{b} \times \hat{\mathbf{n}}_{\perp 1}$. For a RHCP wave the $\perp 1$ component is ahead by 90 degrees in the phase angle. In the inflow region the magnetic field is predominantly in the $x$ direction and the two perpendicular directions correspond to the two coordinate axes as shown.}
\label{figure_rhcp}
\end{figure}

The nature of the waves in the inflow and outflow is further clarified by analyzing the angle between the two perpendicular  fluctuations of the magnetic and electric field. We define the average mean magnetic field as the temporal average of the signal over all the 24 probes along $z$ located at the same $x-y$ location. We then consider the Fourier transform of the two perpendicular directions of the magnetic (Fig.~\ref{figure_F_angle} top) and electric field (Fig.~\ref{figure_F_angle} bottom). To facilitate identifying the polarization of the fluctuations we define the two perpendicular direction as shown in Fig.~\ref{figure_rhcp}.

If the signal is rotating in the same sense of the electrons (i.e. it is right-handed polarized) the $\hat{\mathbf{n}}_{\perp 1}$ signal precedes the $\hat{\mathbf{n}}_{\perp 2}$ signal by 90 degrees in the phase angle. In Fourier space this can be measured as the angle between the complex values of the two Fourier  transforms. This value is obtained taking the ratio of the two transforms and measuring the angle of the complex number in circular coordinates: $\left (\widetilde{B}_{\perp 1}/\widetilde{B}_{\perp 2} \right)_\angle$.

As can be observed in the in the inflow the magnetic perturbation has a well defined +90 degrees angle of polarization in the broad frequency range around the cyclotron region. Only at  very high frequency, where likely the particle noise plays an important role, the polarization is lost.

A polarization angle of 90 degrees in the magnetic field perturbation is not a coincidence, it is an indication of magnetic structures that are frozen into the electrons and rotate in the electron sense. An example of such process is the whistler wave that is right handed circularly polarized. However,  the electric field fluctuations have no clear polarization in this same frequency range, complicating a simple interpretation of the observed processes as whistler waves. The electric field  include both electromagnetic and electrostatic processes, resulting in the absence of a clear polarization. We cannot distinguish in the electric field the contribution from the electromagnetic from the electrostatic field while in the magnetic field of course perturbations can only be electromagnetic.

In the outflow the 90 degree polarization is a fading memory and the angle between $ \widetilde{B}_{\perp 1}$ and $\widetilde{B}_{\perp 2} $ is much broader. Another indication that in the outflow turbulence is becoming fully developed and isotropic. At the highest frequency a new polarization of  $\pm 180$ degrees appears but with a broad spread: the two perpendicular directions of the magnetic field become out of phase, consistent with a linear polarization.  At these very high frequency, however, the results are polluted by particle noise typical of PIC simulations.

\section{Contributions of turbulence to momentum exchange}
\label{ohm}

A key consequence of the two distinctly different regimes of fluctuation in the inflow and outflow is the impact turbulence has on the momentum exchange as measured by the generalized Ohm's law. The electromagnetic-dominated regime does not  produce  sizable anomalous terms because it is not connected with fluctuations in the plasma species. Conversely, the fluctuation regime in the outflow  leads to strong anomalous effects.

The study of  momentum exchange  is based on the generalized Ohm's law that is in essence a rewriting of the electron momentum equation in a two fluid approach \cite{braginskii}. For each term, we separate the contribution from the average and fluctuating part of the fields and the moments. For example, the electric field is decomposed as $\bfE= \langle\bfE\rangle +\delta \bfE$. Substituting the decompositions for each quantity into the generalized Ohm law and  averaging it along $z$, one obtains \cite{braginskii}:
\begin{equation}
e \langle n \rangle \langle \bfE \rangle = m_e \langle n\rangle \frac{d\langle\bfV_e\rangle}{dt}
+ \nabla \cdot  \langle P_e \rangle -\langle \bfJ_e \rangle \times \langle \bfB \rangle  +\Xi,
\end{equation}
where the anomalous terms $\Xi$ are:
\begin{equation}
\Xi = - e \langle \delta n \delta \bfE \rangle + m_e \langle \delta n \frac{d\delta\bfV_e}{dt} \rangle -\langle \delta \bfJ_e  \times \delta \bfB \rangle  ,
\end{equation}
where the three fluctuation-supported terms are the electrostatic (aka anomalous resistivity), inertia and electromagnetic (aka anomalous viscosity~\cite{price2017turbulence}) terms. The fluctuations of the pressure tensor in this formalism do not produce anomalous effects.

\begin{figure}[htbp]
\includegraphics[width=\columnwidth]{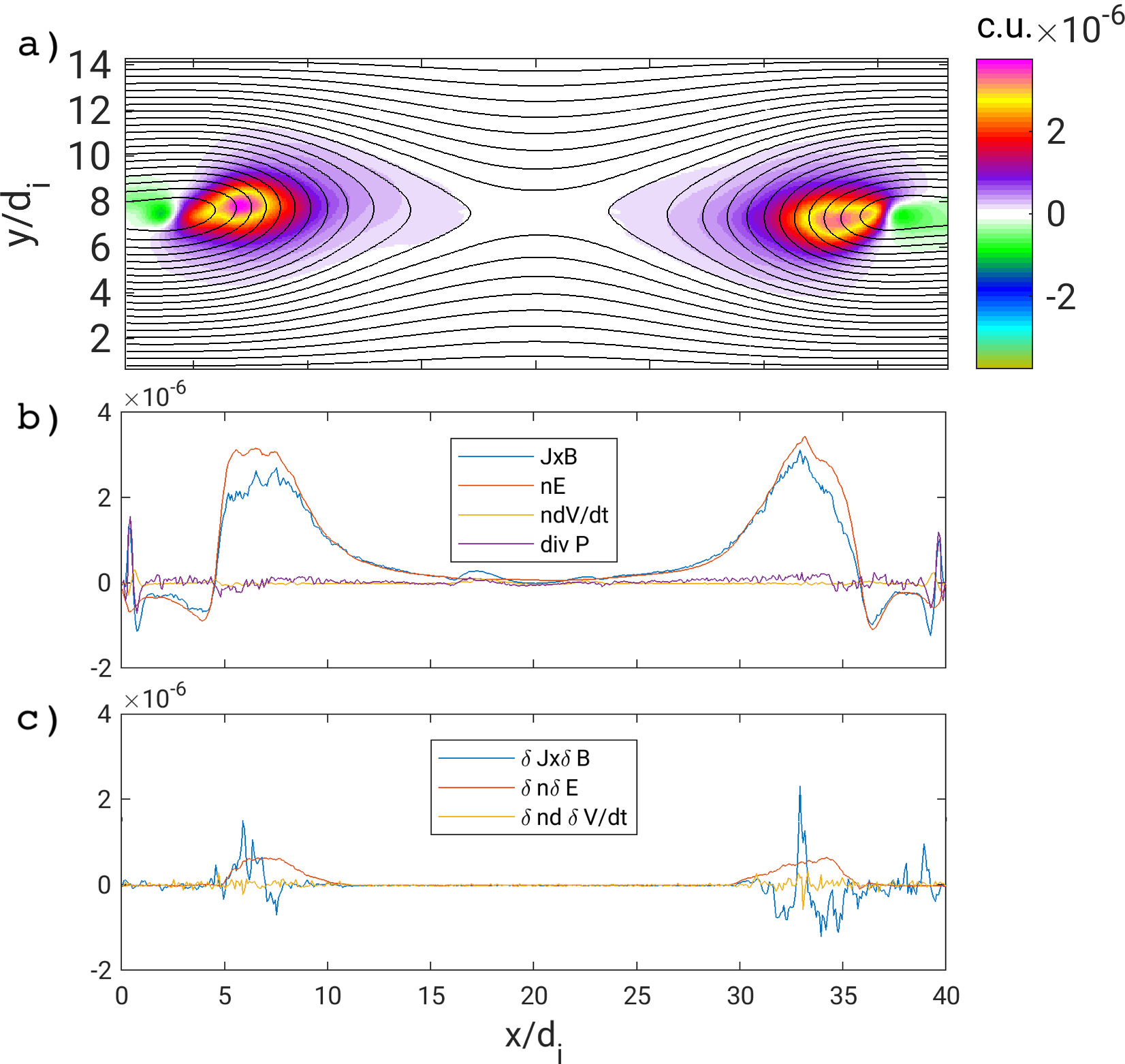}
\caption{False color representation of the $z$-component of the generalized Ohm's law, $e\langle n \rangle \langle E_z \rangle c^2 /m_i \omega_{pi}^4$ (top). Cut along the $x$ axis (at $y=L_y/2$) of the mean  (middle)  and fluctuation (bottom) terms of  the generalized Ohm's law. }
\label{figure_ohm_horizontal}
\end{figure}

\begin{figure}[htbp]
\includegraphics[width=\columnwidth]{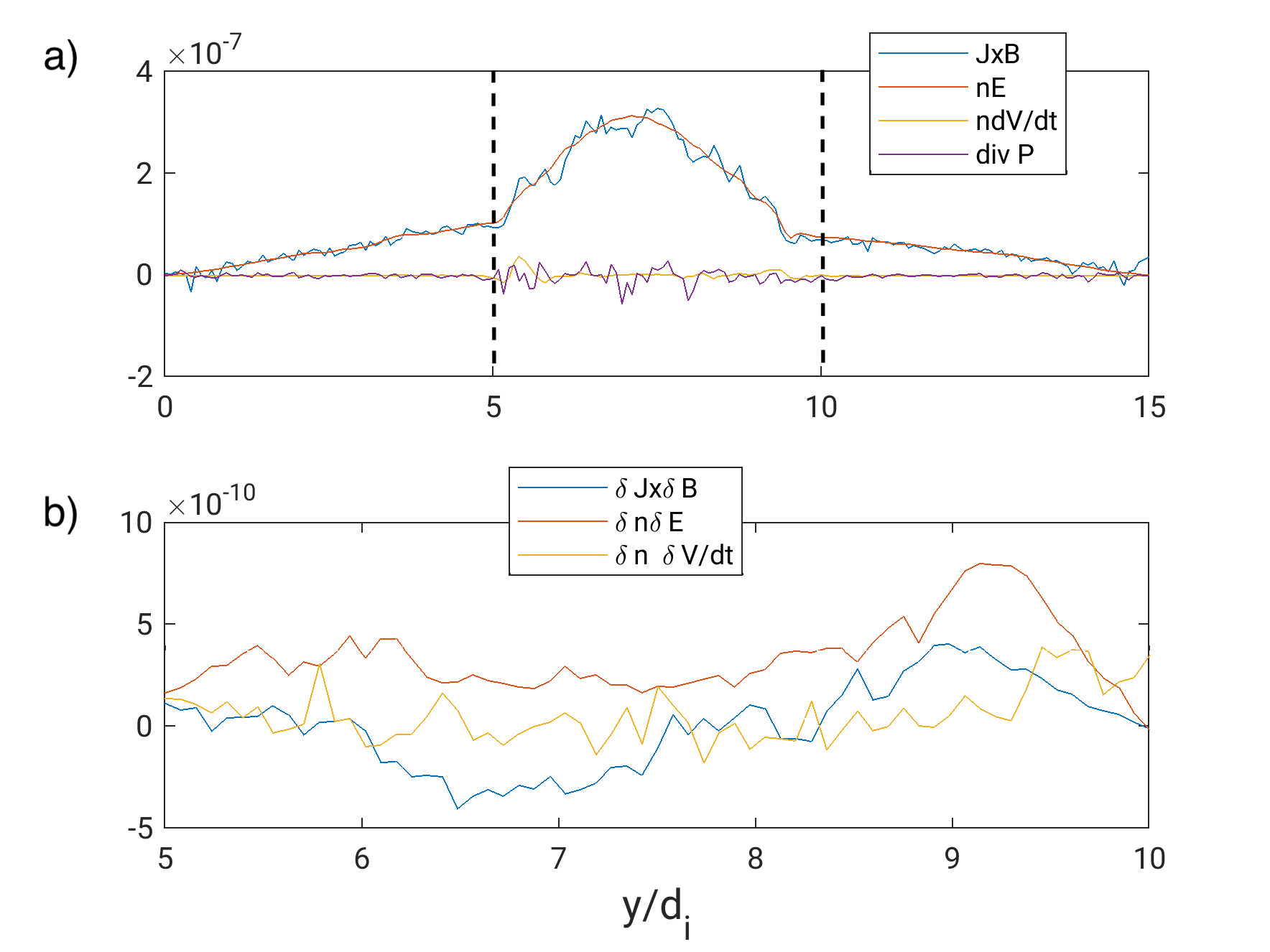}
\caption{ Cut along the $y$ axis (at $x=L_x/2$)  of the terms of the generalized Ohm's law. Panel a shows the mean terms. Panel b reports the contributions of the fluctuations are reported for the central region delimited by the vertical dashed lines in panel a.}
\label{figure_ohm_vertical}
\end{figure}

Figure \ref{figure_ohm_horizontal} and \ref{figure_ohm_vertical} show the different terms in the generalized Ohm's law for the $z$-component (similar conclusions apply for the other two components, not reported because the main driver of reconnection in the present  case is $E_z$). At this relatively late stage of reconnection, the reconnection electric field has moved with the outgoing fronts and is no longer peaked in the center \cite{wan2008evolutions,sitnov2009dipolarization}. As predicted above, in the electromagnetic-dominated region, the fluctuations are not providing any momentum exchange, the mean and  fluctuation terms remain negligible. In the outflow, instead, strong anomalous effect become relevant.

In Fig.~\ref{figure_ohm_horizontal}, the top panel shows the mean reconnection electric field $\langle E_z \rangle$. The second and third panel report the cut along the $x$ axis (at $y=L_y/2$) of the  terms of the generalized Ohm's law.  Figure \ref{figure_ohm_vertical} reports the corresponding cuts along the $y$ axis (at $x=L_x/2$).
   
The cuts along the $x$ axis pass through the outflow region at $\Phi>0$ and highlights the contributions from the fluctuations in the high beta turbulence region where the particles contribute strongly to the turbulence processes.  The majority of the out of plane electric field is caused by the Hall and advection  terms $\langle J_e\rangle \times \langle \bfB\rangle$, with the pressure term being small in comparison. The anomalous terms, instead, provide a substantial minority contribution, especially the electrostatic fluctuation term, $\delta n \delta E$ but also the electromagnetic term, $\delta J_{e} \times \delta \bfB$. The inertia term is negligible by virtue of the mass ratio of the electrons being very small in the present simulation, a realistic effect because in reality it is even smaller (1836 instead of 256 for hydrogen plasma). 

The cuts along the $y$ axis passes through the inflow region at $\Phi<0$ and highlights the contributions from the fluctuations in the low beta turbulence region where the fluctuations remain primarily concentrated only in the electromagnetic field. Here the out of plane electric field is essentially given only by the $\langle J_e\rangle \times \langle \bfB\rangle$ term: in this region the Ohm's law essentially reduces to the statement of  the electron frozen-in condition: $\langle \bfE \rangle= - \langle \bfv_e \rangle \times \langle \bfB \rangle$. The contribution from the fluctuations is measurable in the simulation but nearly three orders of magnitude smaller than the mean electron frozen-in term, a small fraction of even the mean inertia and pressure tensor terms. Momentum in this regime remains laminar.

\section{Summary and Conclusions}
\label{summary}
In summary, the reported work analyses the fluctuations around a reconnection site using full 3D PIC simulations with electrons and ions both treated as  superparticles. The resolution includes the electron effects down to a fraction of the electron skin depth and electron cyclotron frequency, well beyond the scale of interest of the fluctuations studied. The results can then be considered well converged, as proven by simulations with less resolution but substantially equivalent results not reported here. The main conclusion is that fluctuations in the outflow present the classic scenario of turbulent reconnection with plasma species and electromagnetic fields fluctuating in sync. The inflow and separatrix region, instead, present a different type of fluctuations: the fields fluctuate but the plasma species remain essentially laminar. 

In the inflow, the plasma species pass quickly,  waves are generated by their passage but the cause that created them is quickly removed. The electrons population is rapidly and locally drifting, causing instabilities due to the relative ion-electron drift (Buneman instability~\cite{Goldman2008,Divin2012}) and between different electron populations~\cite{Goldman2014}. Currents associated with the strong drift cause also tearing and Kelvin-Helmholtz instabilities \cite{Daughton:2011,Divin2012, lapenta2014separatrices,fermo2012secondary}. The waves generated at the separatrix layer travel in the inflow region, as is the case for example of Cerenkov emission of whistler waves from electron holes~\cite{Goldman2014} or Kinetic \alf waves produced by the fast kinetic reconnection process and  traveling at super\alf speeds \cite{shay2011super}. The particles travel through the separatrices at high speed but the fluctuating electromagnetic fields in the region linger contributing to the electromagnetic fluctuations observed. 

In the outflow, instead, the species slow down and interact with the ambient plasma forming a pile up region. The conditions arise for the excitation of the lower hybrid drift instability due to the strong density gradients \cite{divin2015evolution,lapenta2015secondary,lapenta2018nonlinear} and interchange-type instabilities caused by the unfavorable curvature of the field lines \cite{guzdar2010simple}. In these regions the plasma cannot escape as it did in the inflow separatrices and  turbulence develops fully, involving both fields and particles \cite{innocenti2016study,lapenta2016energy,pucci2017properties,pucci2018generation,lapenta2018nonlinear}.

The consequence of these two regimes is that the fluctuations produce strong momentum and energy exchange (anomalous viscosity and resistivity) only in the outflow. In the inflow and separatrix regions there is strong but laminar energy  exchange with essentially no anomalous momentum exchange despite the electromagnetic fluctuations present. 



\section*{Acknowledgments}
Giovanni Lapenta thanks Christos Tsironis for the discussions on the nature of the waves observed in the Fourier spectra of the virtual probes. 
This project has received funding from the European Union's Horizon 2020 research and innovation programme under grant agreement No. 776262 (AIDA) and from the NASA grant 80NSSC19K0841. 
This
research used resources of the National Energy Research
Scientific Computing Center, which is supported by the Office
of Science of the US Department of Energy under Contract No.
DE-AC02-05CH11231. Additional computing has been provided
by NASA NAS and NCCS High Performance Computing,
by the Flemish Supercomputing Center (VSC) and by a
PRACE Tier-0 allocation. 
 The work by F. Pucci has been supported by Fonds Wetenschappelijk Onderzoek - Vlaanderen (FWO) through the postdoctoral fellowship 12X0319N.


\begin{thebibliography}{72}
\expandafter\ifx\csname natexlab\endcsname\relax\def\natexlab#1{#1}\fi
\expandafter\ifx\csname bibnamefont\endcsname\relax
  \def\bibnamefont#1{#1}\fi
\expandafter\ifx\csname bibfnamefont\endcsname\relax
  \def\bibfnamefont#1{#1}\fi
\expandafter\ifx\csname citenamefont\endcsname\relax
  \def\citenamefont#1{#1}\fi
\expandafter\ifx\csname url\endcsname\relax
  \def\url#1{\texttt{#1}}\fi
\expandafter\ifx\csname urlprefix\endcsname\relax\def\urlprefix{URL }\fi
\providecommand{\bibinfo}[2]{#2}
\providecommand{\eprint}[2][]{\url{#2}}

\bibitem[{\citenamefont{Zweibel and Yamada}(2016)}]{zweibel2016perspectives}
\bibinfo{author}{\bibfnamefont{E.~G.} \bibnamefont{Zweibel}} \bibnamefont{and}
  \bibinfo{author}{\bibfnamefont{M.}~\bibnamefont{Yamada}},
  \bibinfo{journal}{Proceedings of the Royal Society A: Mathematical, Physical
  and Engineering Sciences} \textbf{\bibinfo{volume}{472}},
  \bibinfo{pages}{20160479} (\bibinfo{year}{2016}).

\bibitem[{\citenamefont{Wesson}(1990)}]{wesson1990sawtooth}
\bibinfo{author}{\bibfnamefont{J.~A.} \bibnamefont{Wesson}},
  \bibinfo{journal}{Nuclear Fusion} \textbf{\bibinfo{volume}{30}},
  \bibinfo{pages}{2545} (\bibinfo{year}{1990}).

\bibitem[{\citenamefont{Sironi and Spitkovsky}(2014)}]{sironi2014relativistic}
\bibinfo{author}{\bibfnamefont{L.}~\bibnamefont{Sironi}} \bibnamefont{and}
  \bibinfo{author}{\bibfnamefont{A.}~\bibnamefont{Spitkovsky}},
  \bibinfo{journal}{The Astrophysical Journal Letters}
  \textbf{\bibinfo{volume}{783}}, \bibinfo{pages}{L21} (\bibinfo{year}{2014}).

\bibitem[{\citenamefont{Wyper et~al.}(2017)\citenamefont{Wyper, Antiochos, and
  DeVore}}]{wyper2017universal}
\bibinfo{author}{\bibfnamefont{P.~F.} \bibnamefont{Wyper}},
  \bibinfo{author}{\bibfnamefont{S.~K.} \bibnamefont{Antiochos}},
  \bibnamefont{and} \bibinfo{author}{\bibfnamefont{C.~R.}
  \bibnamefont{DeVore}}, \bibinfo{journal}{Nature}
  \textbf{\bibinfo{volume}{544}}, \bibinfo{pages}{452} (\bibinfo{year}{2017}).

\bibitem[{\citenamefont{Birn et~al.}(2001)\citenamefont{Birn, Drake, Shay,
  Rogers, Denton, Hesse, Kuznetsova, Ma, Bhattacharjee, Otto
  et~al.}}]{birn2001geospace}
\bibinfo{author}{\bibfnamefont{J.}~\bibnamefont{Birn}},
  \bibinfo{author}{\bibfnamefont{J.}~\bibnamefont{Drake}},
  \bibinfo{author}{\bibfnamefont{M.}~\bibnamefont{Shay}},
  \bibinfo{author}{\bibfnamefont{B.}~\bibnamefont{Rogers}},
  \bibinfo{author}{\bibfnamefont{R.}~\bibnamefont{Denton}},
  \bibinfo{author}{\bibfnamefont{M.}~\bibnamefont{Hesse}},
  \bibinfo{author}{\bibfnamefont{M.}~\bibnamefont{Kuznetsova}},
  \bibinfo{author}{\bibfnamefont{Z.}~\bibnamefont{Ma}},
  \bibinfo{author}{\bibfnamefont{A.}~\bibnamefont{Bhattacharjee}},
  \bibinfo{author}{\bibfnamefont{A.}~\bibnamefont{Otto}}, \bibnamefont{et~al.},
  \bibinfo{journal}{Journal of Geophysical Research: Space Physics}
  \textbf{\bibinfo{volume}{106}}, \bibinfo{pages}{3715} (\bibinfo{year}{2001}).

\bibitem[{\citenamefont{Yamada et~al.}(2010)\citenamefont{Yamada, Kulsrud, and
  Ji}}]{yamada2010magnetic}
\bibinfo{author}{\bibfnamefont{M.}~\bibnamefont{Yamada}},
  \bibinfo{author}{\bibfnamefont{R.}~\bibnamefont{Kulsrud}}, \bibnamefont{and}
  \bibinfo{author}{\bibfnamefont{H.}~\bibnamefont{Ji}},
  \bibinfo{journal}{Reviews of Modern Physics} \textbf{\bibinfo{volume}{82}},
  \bibinfo{pages}{603} (\bibinfo{year}{2010}).

\bibitem[{\citenamefont{Burch et~al.}(2016)\citenamefont{Burch, Torbert, Phan,
  Chen, Moore, Ergun, Eastwood, Gershman, Cassak, Argall
  et~al.}}]{burch2016electron}
\bibinfo{author}{\bibfnamefont{J.}~\bibnamefont{Burch}},
  \bibinfo{author}{\bibfnamefont{R.}~\bibnamefont{Torbert}},
  \bibinfo{author}{\bibfnamefont{T.}~\bibnamefont{Phan}},
  \bibinfo{author}{\bibfnamefont{L.-J.} \bibnamefont{Chen}},
  \bibinfo{author}{\bibfnamefont{T.}~\bibnamefont{Moore}},
  \bibinfo{author}{\bibfnamefont{R.}~\bibnamefont{Ergun}},
  \bibinfo{author}{\bibfnamefont{J.}~\bibnamefont{Eastwood}},
  \bibinfo{author}{\bibfnamefont{D.}~\bibnamefont{Gershman}},
  \bibinfo{author}{\bibfnamefont{P.}~\bibnamefont{Cassak}},
  \bibinfo{author}{\bibfnamefont{M.}~\bibnamefont{Argall}},
  \bibnamefont{et~al.}, \bibinfo{journal}{Science}
  \textbf{\bibinfo{volume}{352}}, \bibinfo{pages}{aaf2939}
  (\bibinfo{year}{2016}).

\bibitem[{\citenamefont{Ergun et~al.}(2016)\citenamefont{Ergun, Goodrich,
  Wilder, Holmes, Stawarz, Eriksson, Sturner, Malaspina, Usanova, Torbert
  et~al.}}]{ergun2016magnetospheric}
\bibinfo{author}{\bibfnamefont{R.}~\bibnamefont{Ergun}},
  \bibinfo{author}{\bibfnamefont{K.}~\bibnamefont{Goodrich}},
  \bibinfo{author}{\bibfnamefont{F.}~\bibnamefont{Wilder}},
  \bibinfo{author}{\bibfnamefont{J.}~\bibnamefont{Holmes}},
  \bibinfo{author}{\bibfnamefont{J.}~\bibnamefont{Stawarz}},
  \bibinfo{author}{\bibfnamefont{S.}~\bibnamefont{Eriksson}},
  \bibinfo{author}{\bibfnamefont{A.}~\bibnamefont{Sturner}},
  \bibinfo{author}{\bibfnamefont{D.}~\bibnamefont{Malaspina}},
  \bibinfo{author}{\bibfnamefont{M.}~\bibnamefont{Usanova}},
  \bibinfo{author}{\bibfnamefont{R.}~\bibnamefont{Torbert}},
  \bibnamefont{et~al.}, \bibinfo{journal}{Physical review letters}
  \textbf{\bibinfo{volume}{116}}, \bibinfo{pages}{235102}
  (\bibinfo{year}{2016}).

\bibitem[{\citenamefont{Ergun et~al.}(2017)\citenamefont{Ergun, Chen, Wilder,
  Ahmadi, Eriksson, Usanova, Goodrich, Holmes, Sturner, Malaspina
  et~al.}}]{ergun2017drift}
\bibinfo{author}{\bibfnamefont{R.}~\bibnamefont{Ergun}},
  \bibinfo{author}{\bibfnamefont{L.-J.} \bibnamefont{Chen}},
  \bibinfo{author}{\bibfnamefont{F.}~\bibnamefont{Wilder}},
  \bibinfo{author}{\bibfnamefont{N.}~\bibnamefont{Ahmadi}},
  \bibinfo{author}{\bibfnamefont{S.}~\bibnamefont{Eriksson}},
  \bibinfo{author}{\bibfnamefont{M.}~\bibnamefont{Usanova}},
  \bibinfo{author}{\bibfnamefont{K.}~\bibnamefont{Goodrich}},
  \bibinfo{author}{\bibfnamefont{J.}~\bibnamefont{Holmes}},
  \bibinfo{author}{\bibfnamefont{A.}~\bibnamefont{Sturner}},
  \bibinfo{author}{\bibfnamefont{D.}~\bibnamefont{Malaspina}},
  \bibnamefont{et~al.}, \bibinfo{journal}{Geophysical Research Letters}
  \textbf{\bibinfo{volume}{44}}, \bibinfo{pages}{2978} (\bibinfo{year}{2017}).

\bibitem[{\citenamefont{Retin{\`o} et~al.}(2007)\citenamefont{Retin{\`o},
  Sundkvist, Vaivads, Mozer, Andr{\'e}, and Owen}}]{retino2007situ}
\bibinfo{author}{\bibfnamefont{A.}~\bibnamefont{Retin{\`o}}},
  \bibinfo{author}{\bibfnamefont{D.}~\bibnamefont{Sundkvist}},
  \bibinfo{author}{\bibfnamefont{A.}~\bibnamefont{Vaivads}},
  \bibinfo{author}{\bibfnamefont{F.}~\bibnamefont{Mozer}},
  \bibinfo{author}{\bibfnamefont{M.}~\bibnamefont{Andr{\'e}}},
  \bibnamefont{and} \bibinfo{author}{\bibfnamefont{C.}~\bibnamefont{Owen}},
  \bibinfo{journal}{Nature Physics} \textbf{\bibinfo{volume}{3}},
  \bibinfo{pages}{236} (\bibinfo{year}{2007}).

\bibitem[{\citenamefont{Eastwood et~al.}(2009)\citenamefont{Eastwood, Phan,
  Bale, and Tjulin}}]{eastwood2009observations}
\bibinfo{author}{\bibfnamefont{J.}~\bibnamefont{Eastwood}},
  \bibinfo{author}{\bibfnamefont{T.}~\bibnamefont{Phan}},
  \bibinfo{author}{\bibfnamefont{S.}~\bibnamefont{Bale}}, \bibnamefont{and}
  \bibinfo{author}{\bibfnamefont{A.}~\bibnamefont{Tjulin}},
  \bibinfo{journal}{Physical review letters} \textbf{\bibinfo{volume}{102}},
  \bibinfo{pages}{035001} (\bibinfo{year}{2009}).

\bibitem[{\citenamefont{Ji et~al.}(2004)\citenamefont{Ji, Terry, Yamada,
  Kulsrud, Kuritsyn, and Ren}}]{ji2004electromagnetic}
\bibinfo{author}{\bibfnamefont{H.}~\bibnamefont{Ji}},
  \bibinfo{author}{\bibfnamefont{S.}~\bibnamefont{Terry}},
  \bibinfo{author}{\bibfnamefont{M.}~\bibnamefont{Yamada}},
  \bibinfo{author}{\bibfnamefont{R.}~\bibnamefont{Kulsrud}},
  \bibinfo{author}{\bibfnamefont{A.}~\bibnamefont{Kuritsyn}}, \bibnamefont{and}
  \bibinfo{author}{\bibfnamefont{Y.}~\bibnamefont{Ren}},
  \bibinfo{journal}{Physical review letters} \textbf{\bibinfo{volume}{92}},
  \bibinfo{pages}{115001} (\bibinfo{year}{2004}).

\bibitem[{\citenamefont{Huba et~al.}(1977)\citenamefont{Huba, Gladd, and
  Papadopoulos}}]{huba1977lower}
\bibinfo{author}{\bibfnamefont{J.}~\bibnamefont{Huba}},
  \bibinfo{author}{\bibfnamefont{N.}~\bibnamefont{Gladd}}, \bibnamefont{and}
  \bibinfo{author}{\bibfnamefont{K.}~\bibnamefont{Papadopoulos}},
  \bibinfo{journal}{Geophysical Research Letters} \textbf{\bibinfo{volume}{4}},
  \bibinfo{pages}{125} (\bibinfo{year}{1977}).

\bibitem[{\citenamefont{Vaivads et~al.}(2004)\citenamefont{Vaivads, Andr{\'e},
  Buchert, Wahlund, Fazakerley, and Cornilleau-Wehrlin}}]{vaivads2004cluster}
\bibinfo{author}{\bibfnamefont{A.}~\bibnamefont{Vaivads}},
  \bibinfo{author}{\bibfnamefont{M.}~\bibnamefont{Andr{\'e}}},
  \bibinfo{author}{\bibfnamefont{S.}~\bibnamefont{Buchert}},
  \bibinfo{author}{\bibfnamefont{J.-E.} \bibnamefont{Wahlund}},
  \bibinfo{author}{\bibfnamefont{A.}~\bibnamefont{Fazakerley}},
  \bibnamefont{and}
  \bibinfo{author}{\bibfnamefont{N.}~\bibnamefont{Cornilleau-Wehrlin}},
  \bibinfo{journal}{Geophysical research letters}
  \textbf{\bibinfo{volume}{31}}, \bibinfo{pages}{L03803}
  (\bibinfo{year}{2004}).

\bibitem[{\citenamefont{Carter et~al.}(2001)\citenamefont{Carter, Ji,
  Trintchouk, Yamada, and Kulsrud}}]{carter2001measurement}
\bibinfo{author}{\bibfnamefont{T.}~\bibnamefont{Carter}},
  \bibinfo{author}{\bibfnamefont{H.}~\bibnamefont{Ji}},
  \bibinfo{author}{\bibfnamefont{F.}~\bibnamefont{Trintchouk}},
  \bibinfo{author}{\bibfnamefont{M.}~\bibnamefont{Yamada}}, \bibnamefont{and}
  \bibinfo{author}{\bibfnamefont{R.}~\bibnamefont{Kulsrud}},
  \bibinfo{journal}{Physical review letters} \textbf{\bibinfo{volume}{88}},
  \bibinfo{pages}{015001} (\bibinfo{year}{2001}).

\bibitem[{\citenamefont{Norgren et~al.}(2012)\citenamefont{Norgren, Vaivads,
  Khotyaintsev, and Andr{\'e}}}]{norgren2012lower}
\bibinfo{author}{\bibfnamefont{C.}~\bibnamefont{Norgren}},
  \bibinfo{author}{\bibfnamefont{A.}~\bibnamefont{Vaivads}},
  \bibinfo{author}{\bibfnamefont{Y.~V.} \bibnamefont{Khotyaintsev}},
  \bibnamefont{and}
  \bibinfo{author}{\bibfnamefont{M.}~\bibnamefont{Andr{\'e}}},
  \bibinfo{journal}{Physical review letters} \textbf{\bibinfo{volume}{109}},
  \bibinfo{pages}{055001} (\bibinfo{year}{2012}).

\bibitem[{\citenamefont{Karimabadi et~al.}(2013)\citenamefont{Karimabadi,
  Roytershteyn, Daughton, and Liu}}]{karimabadi2013recent}
\bibinfo{author}{\bibfnamefont{H.}~\bibnamefont{Karimabadi}},
  \bibinfo{author}{\bibfnamefont{V.}~\bibnamefont{Roytershteyn}},
  \bibinfo{author}{\bibfnamefont{W.}~\bibnamefont{Daughton}}, \bibnamefont{and}
  \bibinfo{author}{\bibfnamefont{Y.-H.} \bibnamefont{Liu}}, in
  \emph{\bibinfo{booktitle}{Microphysics of Cosmic Plasmas}}
  (\bibinfo{publisher}{Springer}, \bibinfo{year}{2013}), pp.
  \bibinfo{pages}{231--247}.

\bibitem[{\citenamefont{Galeev and Zelenyi}(1976)}]{galeev1976tearing}
\bibinfo{author}{\bibfnamefont{A.}~\bibnamefont{Galeev}} \bibnamefont{and}
  \bibinfo{author}{\bibfnamefont{L.}~\bibnamefont{Zelenyi}},
  \bibinfo{journal}{Zhurnal Eksperimental'noi i Teoreticheskoi Fiziki}
  \textbf{\bibinfo{volume}{70}}, \bibinfo{pages}{2133} (\bibinfo{year}{1976}).

\bibitem[{\citenamefont{Loureiro et~al.}(2007)\citenamefont{Loureiro,
  Schekochihin, and Cowley}}]{loureiro2007instability}
\bibinfo{author}{\bibfnamefont{N.}~\bibnamefont{Loureiro}},
  \bibinfo{author}{\bibfnamefont{A.}~\bibnamefont{Schekochihin}},
  \bibnamefont{and} \bibinfo{author}{\bibfnamefont{S.}~\bibnamefont{Cowley}},
  \bibinfo{journal}{Physics of Plasmas} \textbf{\bibinfo{volume}{14}},
  \bibinfo{pages}{100703} (\bibinfo{year}{2007}).

\bibitem[{\citenamefont{Cattell et~al.}(2005)\citenamefont{Cattell, Dombeck,
  Wygant, Drake, Swisdak, Goldstein, Keith, Fazakerley, Andr{\'e}, Lucek
  et~al.}}]{cattell2005cluster}
\bibinfo{author}{\bibfnamefont{C.}~\bibnamefont{Cattell}},
  \bibinfo{author}{\bibfnamefont{J.}~\bibnamefont{Dombeck}},
  \bibinfo{author}{\bibfnamefont{J.}~\bibnamefont{Wygant}},
  \bibinfo{author}{\bibfnamefont{J.}~\bibnamefont{Drake}},
  \bibinfo{author}{\bibfnamefont{M.}~\bibnamefont{Swisdak}},
  \bibinfo{author}{\bibfnamefont{M.}~\bibnamefont{Goldstein}},
  \bibinfo{author}{\bibfnamefont{W.}~\bibnamefont{Keith}},
  \bibinfo{author}{\bibfnamefont{A.}~\bibnamefont{Fazakerley}},
  \bibinfo{author}{\bibfnamefont{M.}~\bibnamefont{Andr{\'e}}},
  \bibinfo{author}{\bibfnamefont{E.}~\bibnamefont{Lucek}},
  \bibnamefont{et~al.}, \bibinfo{journal}{Journal of Geophysical Research:
  Space Physics} \textbf{\bibinfo{volume}{110}}, \bibinfo{pages}{A01211}
  (\bibinfo{year}{2005}).

\bibitem[{\citenamefont{{Divin} et~al.}(2012)\citenamefont{{Divin}, {Lapenta},
  {Markidis}, {Newman}, and {Goldman}}}]{Divin2012}
\bibinfo{author}{\bibfnamefont{A.}~\bibnamefont{{Divin}}},
  \bibinfo{author}{\bibfnamefont{G.}~\bibnamefont{{Lapenta}}},
  \bibinfo{author}{\bibfnamefont{S.}~\bibnamefont{{Markidis}}},
  \bibinfo{author}{\bibfnamefont{D.~L.} \bibnamefont{{Newman}}},
  \bibnamefont{and} \bibinfo{author}{\bibfnamefont{M.~V.}
  \bibnamefont{{Goldman}}}, \bibinfo{journal}{Physics of Plasmas}
  \textbf{\bibinfo{volume}{19}}, \bibinfo{pages}{042110}
  (\bibinfo{year}{2012}).

\bibitem[{\citenamefont{Fujimoto}(2014)}]{fujimoto2014wave}
\bibinfo{author}{\bibfnamefont{K.}~\bibnamefont{Fujimoto}},
  \bibinfo{journal}{Geophysical Research Letters}
  \textbf{\bibinfo{volume}{41}}, \bibinfo{pages}{2721} (\bibinfo{year}{2014}).

\bibitem[{\citenamefont{Lapenta et~al.}(2014)\citenamefont{Lapenta, Markidis,
  Divin, Newman, and Goldman}}]{lapenta2014separatrices}
\bibinfo{author}{\bibfnamefont{G.}~\bibnamefont{Lapenta}},
  \bibinfo{author}{\bibfnamefont{S.}~\bibnamefont{Markidis}},
  \bibinfo{author}{\bibfnamefont{A.}~\bibnamefont{Divin}},
  \bibinfo{author}{\bibfnamefont{D.}~\bibnamefont{Newman}}, \bibnamefont{and}
  \bibinfo{author}{\bibfnamefont{M.}~\bibnamefont{Goldman}},
  \bibinfo{journal}{Journal of Plasma Physics} pp. \bibinfo{pages}{1--39}
  (\bibinfo{year}{2014}).

\bibitem[{\citenamefont{Scholer et~al.}(2003)\citenamefont{Scholer, Sidorenko,
  Jaroschek, Treumann, and Zeiler}}]{scholer2003onset}
\bibinfo{author}{\bibfnamefont{M.}~\bibnamefont{Scholer}},
  \bibinfo{author}{\bibfnamefont{I.}~\bibnamefont{Sidorenko}},
  \bibinfo{author}{\bibfnamefont{C.}~\bibnamefont{Jaroschek}},
  \bibinfo{author}{\bibfnamefont{R.}~\bibnamefont{Treumann}}, \bibnamefont{and}
  \bibinfo{author}{\bibfnamefont{A.}~\bibnamefont{Zeiler}},
  \bibinfo{journal}{Physics of Plasmas} \textbf{\bibinfo{volume}{10}},
  \bibinfo{pages}{3521} (\bibinfo{year}{2003}).

\bibitem[{\citenamefont{Fujimoto et~al.}(2011)\citenamefont{Fujimoto,
  Shinohara, and Kojima}}]{fujimoto2011reconnection}
\bibinfo{author}{\bibfnamefont{M.}~\bibnamefont{Fujimoto}},
  \bibinfo{author}{\bibfnamefont{I.}~\bibnamefont{Shinohara}},
  \bibnamefont{and} \bibinfo{author}{\bibfnamefont{H.}~\bibnamefont{Kojima}},
  \bibinfo{journal}{Space science reviews} \textbf{\bibinfo{volume}{160}},
  \bibinfo{pages}{123} (\bibinfo{year}{2011}).

\bibitem[{\citenamefont{Yin et~al.}(2008)\citenamefont{Yin, Daughton,
  Karimabadi, Albright, Bowers, and Margulies}}]{yin2008three}
\bibinfo{author}{\bibfnamefont{L.}~\bibnamefont{Yin}},
  \bibinfo{author}{\bibfnamefont{W.}~\bibnamefont{Daughton}},
  \bibinfo{author}{\bibfnamefont{H.}~\bibnamefont{Karimabadi}},
  \bibinfo{author}{\bibfnamefont{B.}~\bibnamefont{Albright}},
  \bibinfo{author}{\bibfnamefont{K.~J.} \bibnamefont{Bowers}},
  \bibnamefont{and}
  \bibinfo{author}{\bibfnamefont{J.}~\bibnamefont{Margulies}},
  \bibinfo{journal}{Physical review letters} \textbf{\bibinfo{volume}{101}},
  \bibinfo{pages}{125001} (\bibinfo{year}{2008}).

\bibitem[{\citenamefont{Divin et~al.}(2015{\natexlab{a}})\citenamefont{Divin,
  Khotyaintsev, Vaivads, Andr{\'e}, Markidis, and
  Lapenta}}]{divin2015evolution}
\bibinfo{author}{\bibfnamefont{A.}~\bibnamefont{Divin}},
  \bibinfo{author}{\bibfnamefont{Y.~V.} \bibnamefont{Khotyaintsev}},
  \bibinfo{author}{\bibfnamefont{A.}~\bibnamefont{Vaivads}},
  \bibinfo{author}{\bibfnamefont{M.}~\bibnamefont{Andr{\'e}}},
  \bibinfo{author}{\bibfnamefont{S.}~\bibnamefont{Markidis}}, \bibnamefont{and}
  \bibinfo{author}{\bibfnamefont{G.}~\bibnamefont{Lapenta}},
  \bibinfo{journal}{Journal of Geophysical Research: Space Physics}
  \textbf{\bibinfo{volume}{120}}, \bibinfo{pages}{2675}
  (\bibinfo{year}{2015}{\natexlab{a}}).

\bibitem[{\citenamefont{Divin et~al.}(2015{\natexlab{b}})\citenamefont{Divin,
  Khotyaintsev, Vaivads, and Andr{\'e}}}]{divin2015lower}
\bibinfo{author}{\bibfnamefont{A.}~\bibnamefont{Divin}},
  \bibinfo{author}{\bibfnamefont{Y.~V.} \bibnamefont{Khotyaintsev}},
  \bibinfo{author}{\bibfnamefont{A.}~\bibnamefont{Vaivads}}, \bibnamefont{and}
  \bibinfo{author}{\bibfnamefont{M.}~\bibnamefont{Andr{\'e}}},
  \bibinfo{journal}{Journal of Geophysical Research: Space Physics}
  (\bibinfo{year}{2015}{\natexlab{b}}).

\bibitem[{\citenamefont{Price et~al.}(2017)\citenamefont{Price, Swisdak, Drake,
  Burch, Cassak, and Ergun}}]{price2017turbulence}
\bibinfo{author}{\bibfnamefont{L.}~\bibnamefont{Price}},
  \bibinfo{author}{\bibfnamefont{M.}~\bibnamefont{Swisdak}},
  \bibinfo{author}{\bibfnamefont{J.~F.} \bibnamefont{Drake}},
  \bibinfo{author}{\bibfnamefont{J.~L.} \bibnamefont{Burch}},
  \bibinfo{author}{\bibfnamefont{P.~A.} \bibnamefont{Cassak}},
  \bibnamefont{and} \bibinfo{author}{\bibfnamefont{R.~E.} \bibnamefont{Ergun}},
  \bibinfo{journal}{Journal of Geophysical Research: Space Physics}
  \textbf{\bibinfo{volume}{122}}, \bibinfo{pages}{11086}
  (\bibinfo{year}{2017}).

\bibitem[{\citenamefont{Le et~al.}(2017)\citenamefont{Le, Daughton, Chen, and
  Egedal}}]{le2017enhanced}
\bibinfo{author}{\bibfnamefont{A.}~\bibnamefont{Le}},
  \bibinfo{author}{\bibfnamefont{W.}~\bibnamefont{Daughton}},
  \bibinfo{author}{\bibfnamefont{L.-J.} \bibnamefont{Chen}}, \bibnamefont{and}
  \bibinfo{author}{\bibfnamefont{J.}~\bibnamefont{Egedal}},
  \bibinfo{journal}{Geophysical Research Letters}
  \textbf{\bibinfo{volume}{44}}, \bibinfo{pages}{2096} (\bibinfo{year}{2017}).

\bibitem[{\citenamefont{Pritchett and Coroniti}(2011)}]{pritchett2011plasma}
\bibinfo{author}{\bibfnamefont{P.~L.} \bibnamefont{Pritchett}}
  \bibnamefont{and} \bibinfo{author}{\bibfnamefont{F.~V.}
  \bibnamefont{Coroniti}}, \bibinfo{journal}{Geophysical Research Letters}
  \textbf{\bibinfo{volume}{38}}, \bibinfo{pages}{L10102}
  (\bibinfo{year}{2011}).

\bibitem[{\citenamefont{Nakamura et~al.}(2016)\citenamefont{Nakamura, Nakamura,
  Baumjohann, Umeda, and Shinohara}}]{nakamura2016three}
\bibinfo{author}{\bibfnamefont{T.~K.~M.} \bibnamefont{Nakamura}},
  \bibinfo{author}{\bibfnamefont{R.}~\bibnamefont{Nakamura}},
  \bibinfo{author}{\bibfnamefont{W.}~\bibnamefont{Baumjohann}},
  \bibinfo{author}{\bibfnamefont{T.}~\bibnamefont{Umeda}}, \bibnamefont{and}
  \bibinfo{author}{\bibfnamefont{I.}~\bibnamefont{Shinohara}},
  \bibinfo{journal}{Geophysical Research Letters}
  \textbf{\bibinfo{volume}{43}}, \bibinfo{pages}{8356} (\bibinfo{year}{2016}).

\bibitem[{\citenamefont{Lapenta and Bettarini}(2011)}]{lapenta2011self}
\bibinfo{author}{\bibfnamefont{G.}~\bibnamefont{Lapenta}} \bibnamefont{and}
  \bibinfo{author}{\bibfnamefont{L.}~\bibnamefont{Bettarini}},
  \bibinfo{journal}{Geophysical Research Letters} \textbf{\bibinfo{volume}{38}}
  (\bibinfo{year}{2011}).

\bibitem[{\citenamefont{Vapirev et~al.}(2013)\citenamefont{Vapirev, Lapenta,
  Divin, Markidis, Henri, Goldman, and Newman}}]{vapirev2013formation}
\bibinfo{author}{\bibfnamefont{A.}~\bibnamefont{Vapirev}},
  \bibinfo{author}{\bibfnamefont{G.}~\bibnamefont{Lapenta}},
  \bibinfo{author}{\bibfnamefont{A.}~\bibnamefont{Divin}},
  \bibinfo{author}{\bibfnamefont{S.}~\bibnamefont{Markidis}},
  \bibinfo{author}{\bibfnamefont{P.}~\bibnamefont{Henri}},
  \bibinfo{author}{\bibfnamefont{M.}~\bibnamefont{Goldman}}, \bibnamefont{and}
  \bibinfo{author}{\bibfnamefont{D.}~\bibnamefont{Newman}},
  \bibinfo{journal}{Journal of Geophysical Research: Space Physics}
  \textbf{\bibinfo{volume}{118}}, \bibinfo{pages}{1435} (\bibinfo{year}{2013}).

\bibitem[{\citenamefont{Sitnov et~al.}(2014)\citenamefont{Sitnov, Merkin,
  Swisdak, Motoba, Buzulukova, Moore, Mauk, and Ohtani}}]{sitnov2014magnetic}
\bibinfo{author}{\bibfnamefont{M.}~\bibnamefont{Sitnov}},
  \bibinfo{author}{\bibfnamefont{V.}~\bibnamefont{Merkin}},
  \bibinfo{author}{\bibfnamefont{M.}~\bibnamefont{Swisdak}},
  \bibinfo{author}{\bibfnamefont{T.}~\bibnamefont{Motoba}},
  \bibinfo{author}{\bibfnamefont{N.}~\bibnamefont{Buzulukova}},
  \bibinfo{author}{\bibfnamefont{T.}~\bibnamefont{Moore}},
  \bibinfo{author}{\bibfnamefont{B.}~\bibnamefont{Mauk}}, \bibnamefont{and}
  \bibinfo{author}{\bibfnamefont{S.}~\bibnamefont{Ohtani}},
  \bibinfo{journal}{Journal of Geophysical Research: Space Physics}
  \textbf{\bibinfo{volume}{119}}, \bibinfo{pages}{7151} (\bibinfo{year}{2014}).

\bibitem[{\citenamefont{Che et~al.}(2011)\citenamefont{Che, Drake, and
  Swisdak}}]{che2011current}
\bibinfo{author}{\bibfnamefont{H.}~\bibnamefont{Che}},
  \bibinfo{author}{\bibfnamefont{J.}~\bibnamefont{Drake}}, \bibnamefont{and}
  \bibinfo{author}{\bibfnamefont{M.}~\bibnamefont{Swisdak}},
  \bibinfo{journal}{Nature} \textbf{\bibinfo{volume}{474}},
  \bibinfo{pages}{184} (\bibinfo{year}{2011}).

\bibitem[{\citenamefont{{Daughton} et~al.}(2011)\citenamefont{{Daughton},
  {Roytershteyn}, {Karimabadi}, {Yin}, {Albright}, {Bergen}, and
  {Bowers}}}]{Daughton:2011}
\bibinfo{author}{\bibfnamefont{W.}~\bibnamefont{{Daughton}}},
  \bibinfo{author}{\bibfnamefont{V.}~\bibnamefont{{Roytershteyn}}},
  \bibinfo{author}{\bibfnamefont{H.}~\bibnamefont{{Karimabadi}}},
  \bibinfo{author}{\bibfnamefont{L.}~\bibnamefont{{Yin}}},
  \bibinfo{author}{\bibfnamefont{B.~J.} \bibnamefont{{Albright}}},
  \bibinfo{author}{\bibfnamefont{B.}~\bibnamefont{{Bergen}}}, \bibnamefont{and}
  \bibinfo{author}{\bibfnamefont{K.~J.} \bibnamefont{{Bowers}}},
  \bibinfo{journal}{Nature Physics} \textbf{\bibinfo{volume}{7}},
  \bibinfo{pages}{539} (\bibinfo{year}{2011}).

\bibitem[{\citenamefont{Markidis et~al.}(2012)\citenamefont{Markidis, Lapenta,
  Divin, Goldman, Newman, and Andersson}}]{markidis2012three}
\bibinfo{author}{\bibfnamefont{S.}~\bibnamefont{Markidis}},
  \bibinfo{author}{\bibfnamefont{G.}~\bibnamefont{Lapenta}},
  \bibinfo{author}{\bibfnamefont{A.}~\bibnamefont{Divin}},
  \bibinfo{author}{\bibfnamefont{M.}~\bibnamefont{Goldman}},
  \bibinfo{author}{\bibfnamefont{D.}~\bibnamefont{Newman}}, \bibnamefont{and}
  \bibinfo{author}{\bibfnamefont{L.}~\bibnamefont{Andersson}},
  \bibinfo{journal}{Physics of Plasmas (1994-present)}
  \textbf{\bibinfo{volume}{19}}, \bibinfo{pages}{032119}
  (\bibinfo{year}{2012}).

\bibitem[{\citenamefont{Fermo et~al.}(2012)\citenamefont{Fermo, Drake, and
  Swisdak}}]{fermo2012secondary}
\bibinfo{author}{\bibfnamefont{R.}~\bibnamefont{Fermo}},
  \bibinfo{author}{\bibfnamefont{J.}~\bibnamefont{Drake}}, \bibnamefont{and}
  \bibinfo{author}{\bibfnamefont{M.}~\bibnamefont{Swisdak}},
  \bibinfo{journal}{Physical review letters} \textbf{\bibinfo{volume}{108}},
  \bibinfo{pages}{255005} (\bibinfo{year}{2012}).

\bibitem[{\citenamefont{Lapenta et~al.}(2011)\citenamefont{Lapenta, Markidis,
  Divin, Goldman, and Newman}}]{lapenta2011bipolar}
\bibinfo{author}{\bibfnamefont{G.}~\bibnamefont{Lapenta}},
  \bibinfo{author}{\bibfnamefont{S.}~\bibnamefont{Markidis}},
  \bibinfo{author}{\bibfnamefont{A.}~\bibnamefont{Divin}},
  \bibinfo{author}{\bibfnamefont{M.}~\bibnamefont{Goldman}}, \bibnamefont{and}
  \bibinfo{author}{\bibfnamefont{D.}~\bibnamefont{Newman}},
  \bibinfo{journal}{Geophysical Research Letters} \textbf{\bibinfo{volume}{38}}
  (\bibinfo{year}{2011}).

\bibitem[{\citenamefont{{Goldman} et~al.}(2014)\citenamefont{{Goldman},
  {Newman}, {Lapenta}, {Andersson}, {Gosling}, {Eriksson}, {Markidis},
  {Eastwood}, and {Ergun}}}]{Goldman2014}
\bibinfo{author}{\bibfnamefont{M.~V.} \bibnamefont{{Goldman}}},
  \bibinfo{author}{\bibfnamefont{D.~L.} \bibnamefont{{Newman}}},
  \bibinfo{author}{\bibfnamefont{G.}~\bibnamefont{{Lapenta}}},
  \bibinfo{author}{\bibfnamefont{L.}~\bibnamefont{{Andersson}}},
  \bibinfo{author}{\bibfnamefont{J.~T.} \bibnamefont{{Gosling}}},
  \bibinfo{author}{\bibfnamefont{S.}~\bibnamefont{{Eriksson}}},
  \bibinfo{author}{\bibfnamefont{S.}~\bibnamefont{{Markidis}}},
  \bibinfo{author}{\bibfnamefont{J.~P.} \bibnamefont{{Eastwood}}},
  \bibnamefont{and} \bibinfo{author}{\bibfnamefont{R.}~\bibnamefont{{Ergun}}},
  \bibinfo{journal}{Physical Review Letters} \textbf{\bibinfo{volume}{112}},
  \bibinfo{eid}{145002} (\bibinfo{year}{2014}).

\bibitem[{\citenamefont{Gary and Karimabadi}(2006)}]{gary2006linear}
\bibinfo{author}{\bibfnamefont{S.~P.} \bibnamefont{Gary}} \bibnamefont{and}
  \bibinfo{author}{\bibfnamefont{H.}~\bibnamefont{Karimabadi}},
  \bibinfo{journal}{Journal of Geophysical Research: Space Physics}
  \textbf{\bibinfo{volume}{111}} (\bibinfo{year}{2006}).

\bibitem[{\citenamefont{Fujimoto and Sydora}(2008)}]{fujimoto2008whistler}
\bibinfo{author}{\bibfnamefont{K.}~\bibnamefont{Fujimoto}} \bibnamefont{and}
  \bibinfo{author}{\bibfnamefont{R.~D.} \bibnamefont{Sydora}},
  \bibinfo{journal}{Geophysical Research Letters} \textbf{\bibinfo{volume}{35}}
  (\bibinfo{year}{2008}).

\bibitem[{\citenamefont{Markidis et~al.}(2010)\citenamefont{Markidis, Lapenta
  et~al.}}]{markidis2010multi}
\bibinfo{author}{\bibfnamefont{S.}~\bibnamefont{Markidis}},
  \bibinfo{author}{\bibfnamefont{G.}~\bibnamefont{Lapenta}},
  \bibnamefont{et~al.}, \bibinfo{journal}{Mathematics and Computers in
  Simulation} \textbf{\bibinfo{volume}{80}}, \bibinfo{pages}{1509}
  (\bibinfo{year}{2010}).

\bibitem[{\citenamefont{Lapenta et~al.}(2015)\citenamefont{Lapenta, Markidis,
  Goldman, and Newman}}]{lapenta2015secondary}
\bibinfo{author}{\bibfnamefont{G.}~\bibnamefont{Lapenta}},
  \bibinfo{author}{\bibfnamefont{S.}~\bibnamefont{Markidis}},
  \bibinfo{author}{\bibfnamefont{M.~V.} \bibnamefont{Goldman}},
  \bibnamefont{and} \bibinfo{author}{\bibfnamefont{D.~L.}
  \bibnamefont{Newman}}, \bibinfo{journal}{Nature Physics}
  \textbf{\bibinfo{volume}{11}}, \bibinfo{pages}{690} (\bibinfo{year}{2015}).

\bibitem[{\citenamefont{{Harris}}(1962)}]{Harris1962}
\bibinfo{author}{\bibfnamefont{E.~G.} \bibnamefont{{Harris}}},
  \bibinfo{journal}{Il Nuovo Cimento (1955-1965)}
  \textbf{\bibinfo{volume}{23}}, \bibinfo{pages}{115} (\bibinfo{year}{1962}).

\bibitem[{\citenamefont{Runov et~al.}(2006)\citenamefont{Runov, Sergeev,
  Nakamura, Baumjohann, Apatenkov, Asano, Takada, Volwerk, V{\"o}r{\"o}s, Zhang
  et~al.}}]{runov2006local}
\bibinfo{author}{\bibfnamefont{A.}~\bibnamefont{Runov}},
  \bibinfo{author}{\bibfnamefont{V.}~\bibnamefont{Sergeev}},
  \bibinfo{author}{\bibfnamefont{R.}~\bibnamefont{Nakamura}},
  \bibinfo{author}{\bibfnamefont{W.}~\bibnamefont{Baumjohann}},
  \bibinfo{author}{\bibfnamefont{S.}~\bibnamefont{Apatenkov}},
  \bibinfo{author}{\bibfnamefont{Y.}~\bibnamefont{Asano}},
  \bibinfo{author}{\bibfnamefont{T.}~\bibnamefont{Takada}},
  \bibinfo{author}{\bibfnamefont{M.}~\bibnamefont{Volwerk}},
  \bibinfo{author}{\bibfnamefont{Z.}~\bibnamefont{V{\"o}r{\"o}s}},
  \bibinfo{author}{\bibfnamefont{T.}~\bibnamefont{Zhang}},
  \bibnamefont{et~al.}, in \emph{\bibinfo{booktitle}{Annales Geophysicae}}
  (\bibinfo{year}{2006}), vol.~\bibinfo{volume}{24}, pp.
  \bibinfo{pages}{247--262}.

\bibitem[{\citenamefont{Eriksson et~al.}(2014)\citenamefont{Eriksson, Newman,
  Lapenta, and Angelopoulos}}]{eriksson2014signatures}
\bibinfo{author}{\bibfnamefont{S.}~\bibnamefont{Eriksson}},
  \bibinfo{author}{\bibfnamefont{D.}~\bibnamefont{Newman}},
  \bibinfo{author}{\bibfnamefont{G.}~\bibnamefont{Lapenta}}, \bibnamefont{and}
  \bibinfo{author}{\bibfnamefont{V.}~\bibnamefont{Angelopoulos}},
  \bibinfo{journal}{Plasma Physics and Controlled Fusion}
  \textbf{\bibinfo{volume}{56}}, \bibinfo{pages}{064008}
  (\bibinfo{year}{2014}).

\bibitem[{\citenamefont{Pucci et~al.}(2017)\citenamefont{Pucci, Servidio,
  Sorriso-Valvo, Olshevsky, Matthaeus, Malara, Goldman, Newman, and
  Lapenta}}]{pucci2017properties}
\bibinfo{author}{\bibfnamefont{F.}~\bibnamefont{Pucci}},
  \bibinfo{author}{\bibfnamefont{S.}~\bibnamefont{Servidio}},
  \bibinfo{author}{\bibfnamefont{L.}~\bibnamefont{Sorriso-Valvo}},
  \bibinfo{author}{\bibfnamefont{V.}~\bibnamefont{Olshevsky}},
  \bibinfo{author}{\bibfnamefont{W.}~\bibnamefont{Matthaeus}},
  \bibinfo{author}{\bibfnamefont{F.}~\bibnamefont{Malara}},
  \bibinfo{author}{\bibfnamefont{M.}~\bibnamefont{Goldman}},
  \bibinfo{author}{\bibfnamefont{D.}~\bibnamefont{Newman}}, \bibnamefont{and}
  \bibinfo{author}{\bibfnamefont{G.}~\bibnamefont{Lapenta}},
  \bibinfo{journal}{The Astrophysical Journal} \textbf{\bibinfo{volume}{841}},
  \bibinfo{pages}{60} (\bibinfo{year}{2017}).

\bibitem[{\citenamefont{Daughton}(2002)}]{daughton02}
\bibinfo{author}{\bibfnamefont{W.}~\bibnamefont{Daughton}},
  \bibinfo{journal}{Phys. Plasmas} \textbf{\bibinfo{volume}{9}},
  \bibinfo{pages}{3668} (\bibinfo{year}{2002}).

\bibitem[{\citenamefont{Lapenta et~al.}(2003)\citenamefont{Lapenta, Brackbill,
  and Daughton}}]{lapenta2003unexpected}
\bibinfo{author}{\bibfnamefont{G.}~\bibnamefont{Lapenta}},
  \bibinfo{author}{\bibfnamefont{J.}~\bibnamefont{Brackbill}},
  \bibnamefont{and} \bibinfo{author}{\bibfnamefont{W.}~\bibnamefont{Daughton}},
  \bibinfo{journal}{Physics of plasmas} \textbf{\bibinfo{volume}{10}},
  \bibinfo{pages}{1577} (\bibinfo{year}{2003}).

\bibitem[{\citenamefont{Karimabadi et~al.}(2003)\citenamefont{Karimabadi,
  Pritchett, Daughton, and Krauss-Varban}}]{karimabadi2003ion}
\bibinfo{author}{\bibfnamefont{H.}~\bibnamefont{Karimabadi}},
  \bibinfo{author}{\bibfnamefont{P.}~\bibnamefont{Pritchett}},
  \bibinfo{author}{\bibfnamefont{W.}~\bibnamefont{Daughton}}, \bibnamefont{and}
  \bibinfo{author}{\bibfnamefont{D.}~\bibnamefont{Krauss-Varban}},
  \bibinfo{journal}{Journal of Geophysical Research: Space Physics}
  \textbf{\bibinfo{volume}{108}} (\bibinfo{year}{2003}).

\bibitem[{\citenamefont{Aunai et~al.}(2011)\citenamefont{Aunai, Belmont, and
  Smets}}]{aunai2011energy}
\bibinfo{author}{\bibfnamefont{N.}~\bibnamefont{Aunai}},
  \bibinfo{author}{\bibfnamefont{G.}~\bibnamefont{Belmont}}, \bibnamefont{and}
  \bibinfo{author}{\bibfnamefont{R.}~\bibnamefont{Smets}},
  \bibinfo{journal}{Physics of Plasmas} \textbf{\bibinfo{volume}{18}},
  \bibinfo{pages}{122901} (\bibinfo{year}{2011}).

\bibitem[{\citenamefont{Pucci et~al.}(2018)\citenamefont{Pucci, Matthaeus,
  Chasapis, Servidio, Sorriso-Valvo, Olshevsky, Newman, Goldman, and
  Lapenta}}]{pucci2018generation}
\bibinfo{author}{\bibfnamefont{F.}~\bibnamefont{Pucci}},
  \bibinfo{author}{\bibfnamefont{W.~H.} \bibnamefont{Matthaeus}},
  \bibinfo{author}{\bibfnamefont{A.}~\bibnamefont{Chasapis}},
  \bibinfo{author}{\bibfnamefont{S.}~\bibnamefont{Servidio}},
  \bibinfo{author}{\bibfnamefont{L.}~\bibnamefont{Sorriso-Valvo}},
  \bibinfo{author}{\bibfnamefont{V.}~\bibnamefont{Olshevsky}},
  \bibinfo{author}{\bibfnamefont{D.}~\bibnamefont{Newman}},
  \bibinfo{author}{\bibfnamefont{M.}~\bibnamefont{Goldman}}, \bibnamefont{and}
  \bibinfo{author}{\bibfnamefont{G.}~\bibnamefont{Lapenta}},
  \bibinfo{journal}{The Astrophysical Journal} \textbf{\bibinfo{volume}{867}},
  \bibinfo{pages}{10} (\bibinfo{year}{2018}).

\bibitem[{\citenamefont{Swanson}(2003)}]{swanson}
\bibinfo{author}{\bibfnamefont{D.~G.} \bibnamefont{Swanson}},
  \emph{\bibinfo{title}{Plasma Waves; 2nd ed.}}, Plasma physics
  (\bibinfo{publisher}{IOP}, \bibinfo{address}{Bristol}, \bibinfo{year}{2003}).

\bibitem[{\citenamefont{Goodrich et~al.}(2018)\citenamefont{Goodrich, Ergun,
  Schwartz, Wilson~III, Newman, Wilder, Holmes, Johlander, Burch, Torbert
  et~al.}}]{goodrich2018mms}
\bibinfo{author}{\bibfnamefont{K.~A.} \bibnamefont{Goodrich}},
  \bibinfo{author}{\bibfnamefont{R.}~\bibnamefont{Ergun}},
  \bibinfo{author}{\bibfnamefont{S.~J.} \bibnamefont{Schwartz}},
  \bibinfo{author}{\bibfnamefont{L.~B.} \bibnamefont{Wilson~III}},
  \bibinfo{author}{\bibfnamefont{D.}~\bibnamefont{Newman}},
  \bibinfo{author}{\bibfnamefont{F.~D.} \bibnamefont{Wilder}},
  \bibinfo{author}{\bibfnamefont{J.}~\bibnamefont{Holmes}},
  \bibinfo{author}{\bibfnamefont{A.}~\bibnamefont{Johlander}},
  \bibinfo{author}{\bibfnamefont{J.}~\bibnamefont{Burch}},
  \bibinfo{author}{\bibfnamefont{R.}~\bibnamefont{Torbert}},
  \bibnamefont{et~al.}, \bibinfo{journal}{Journal of Geophysical Research:
  Space Physics} \textbf{\bibinfo{volume}{123}}, \bibinfo{pages}{9430}
  (\bibinfo{year}{2018}).

\bibitem[{\citenamefont{Li}(2018)}]{wenya-mms}
\bibinfo{author}{\bibfnamefont{W.}~\bibnamefont{Li}}, in
  \emph{\bibinfo{booktitle}{MMS workshop, Bergen, 12-14 June}}
  (\bibinfo{year}{2018}).

\bibitem[{\citenamefont{Dokgo}(2019)}]{dokgo-mms}
\bibinfo{author}{\bibfnamefont{K.}~\bibnamefont{Dokgo}}, in
  \emph{\bibinfo{booktitle}{fifth Magnetospheric Multiscale (MMS) community
  workshop, Biarritz, 21-24 October}} (\bibinfo{year}{2019}).

\bibitem[{\citenamefont{Burch et~al.}(2019)\citenamefont{Burch, Dokgo, Hwang,
  Torbert, Graham, Webster, Ergun, Giles, Allen, Chen et~al.}}]{burch2019high}
\bibinfo{author}{\bibfnamefont{J.}~\bibnamefont{Burch}},
  \bibinfo{author}{\bibfnamefont{K.}~\bibnamefont{Dokgo}},
  \bibinfo{author}{\bibfnamefont{K.}~\bibnamefont{Hwang}},
  \bibinfo{author}{\bibfnamefont{R.}~\bibnamefont{Torbert}},
  \bibinfo{author}{\bibfnamefont{D.~B.} \bibnamefont{Graham}},
  \bibinfo{author}{\bibfnamefont{J.}~\bibnamefont{Webster}},
  \bibinfo{author}{\bibfnamefont{R.}~\bibnamefont{Ergun}},
  \bibinfo{author}{\bibfnamefont{B.}~\bibnamefont{Giles}},
  \bibinfo{author}{\bibfnamefont{R.}~\bibnamefont{Allen}},
  \bibinfo{author}{\bibfnamefont{L.-J.} \bibnamefont{Chen}},
  \bibnamefont{et~al.}, \bibinfo{journal}{Geophysical Research Letters}
  \textbf{\bibinfo{volume}{46}}, \bibinfo{pages}{4089} (\bibinfo{year}{2019}).

\bibitem[{\citenamefont{Dokgo et~al.}(2019)\citenamefont{Dokgo, Hwang, Burch,
  Choi, Yoon, Sibeck, and Graham}}]{dokgo2019high}
\bibinfo{author}{\bibfnamefont{K.}~\bibnamefont{Dokgo}},
  \bibinfo{author}{\bibfnamefont{K.-J.} \bibnamefont{Hwang}},
  \bibinfo{author}{\bibfnamefont{J.~L.} \bibnamefont{Burch}},
  \bibinfo{author}{\bibfnamefont{E.}~\bibnamefont{Choi}},
  \bibinfo{author}{\bibfnamefont{P.~H.} \bibnamefont{Yoon}},
  \bibinfo{author}{\bibfnamefont{D.~G.} \bibnamefont{Sibeck}},
  \bibnamefont{and} \bibinfo{author}{\bibfnamefont{D.~B.}
  \bibnamefont{Graham}}, \bibinfo{journal}{Geophysical research letters}
  \textbf{\bibinfo{volume}{46}}, \bibinfo{pages}{7873} (\bibinfo{year}{2019}).

\bibitem[{\citenamefont{Huba et~al.}(1980)\citenamefont{Huba, Drake, and
  Gladd}}]{huba1980lower}
\bibinfo{author}{\bibfnamefont{J.}~\bibnamefont{Huba}},
  \bibinfo{author}{\bibfnamefont{J.}~\bibnamefont{Drake}}, \bibnamefont{and}
  \bibinfo{author}{\bibfnamefont{N.}~\bibnamefont{Gladd}},
  \bibinfo{journal}{The Physics of Fluids} \textbf{\bibinfo{volume}{23}},
  \bibinfo{pages}{552} (\bibinfo{year}{1980}).

\bibitem[{\citenamefont{Daughton}(2003)}]{daughton2003electromagnetic}
\bibinfo{author}{\bibfnamefont{W.}~\bibnamefont{Daughton}},
  \bibinfo{journal}{Physics of Plasmas} \textbf{\bibinfo{volume}{10}},
  \bibinfo{pages}{3103} (\bibinfo{year}{2003}).

\bibitem[{\citenamefont{Innocenti et~al.}(2016)\citenamefont{Innocenti,
  Norgren, Newman, Goldman, Markidis, and Lapenta}}]{innocenti2016study}
\bibinfo{author}{\bibfnamefont{M.}~\bibnamefont{Innocenti}},
  \bibinfo{author}{\bibfnamefont{C.}~\bibnamefont{Norgren}},
  \bibinfo{author}{\bibfnamefont{D.}~\bibnamefont{Newman}},
  \bibinfo{author}{\bibfnamefont{M.}~\bibnamefont{Goldman}},
  \bibinfo{author}{\bibfnamefont{S.}~\bibnamefont{Markidis}}, \bibnamefont{and}
  \bibinfo{author}{\bibfnamefont{G.}~\bibnamefont{Lapenta}},
  \bibinfo{journal}{Physics of Plasmas} \textbf{\bibinfo{volume}{23}},
  \bibinfo{pages}{052902} (\bibinfo{year}{2016}).

\bibitem[{\citenamefont{Vapirev et~al.}(2015)\citenamefont{Vapirev, Deca,
  Lapenta, Markidis, Hur, and Cambier}}]{vapirev2015initial}
\bibinfo{author}{\bibfnamefont{A.}~\bibnamefont{Vapirev}},
  \bibinfo{author}{\bibfnamefont{J.}~\bibnamefont{Deca}},
  \bibinfo{author}{\bibfnamefont{G.}~\bibnamefont{Lapenta}},
  \bibinfo{author}{\bibfnamefont{S.}~\bibnamefont{Markidis}},
  \bibinfo{author}{\bibfnamefont{I.}~\bibnamefont{Hur}}, \bibnamefont{and}
  \bibinfo{author}{\bibfnamefont{J.-L.} \bibnamefont{Cambier}},
  \bibinfo{journal}{Concurrency and Computation: Practice and Experience}
  \textbf{\bibinfo{volume}{27}}, \bibinfo{pages}{581} (\bibinfo{year}{2015}).

\bibitem[{\citenamefont{Braginskii}(1965)}]{braginskii}
\bibinfo{author}{\bibfnamefont{S.}~\bibnamefont{Braginskii}},
  \bibinfo{journal}{Rev. Plasma Phys.} \textbf{\bibinfo{volume}{1}},
  \bibinfo{pages}{205} (\bibinfo{year}{1965}).

\bibitem[{\citenamefont{Wan and Lapenta}(2008)}]{wan2008evolutions}
\bibinfo{author}{\bibfnamefont{W.}~\bibnamefont{Wan}} \bibnamefont{and}
  \bibinfo{author}{\bibfnamefont{G.}~\bibnamefont{Lapenta}},
  \bibinfo{journal}{Physics of Plasmas} \textbf{\bibinfo{volume}{15}},
  \bibinfo{pages}{102302} (\bibinfo{year}{2008}).

\bibitem[{\citenamefont{Sitnov et~al.}(2009)\citenamefont{Sitnov, Swisdak, and
  Divin}}]{sitnov2009dipolarization}
\bibinfo{author}{\bibfnamefont{M.}~\bibnamefont{Sitnov}},
  \bibinfo{author}{\bibfnamefont{M.}~\bibnamefont{Swisdak}}, \bibnamefont{and}
  \bibinfo{author}{\bibfnamefont{A.}~\bibnamefont{Divin}},
  \bibinfo{journal}{Journal of Geophysical Research: Space Physics
  (1978--2012)} \textbf{\bibinfo{volume}{114}}, \bibinfo{pages}{L09805}
  (\bibinfo{year}{2009}).

\bibitem[{\citenamefont{{Goldman} et~al.}(2008)\citenamefont{{Goldman},
  {Newman}, and {Pritchett}}}]{Goldman2008}
\bibinfo{author}{\bibfnamefont{M.~V.} \bibnamefont{{Goldman}}},
  \bibinfo{author}{\bibfnamefont{D.~L.} \bibnamefont{{Newman}}},
  \bibnamefont{and}
  \bibinfo{author}{\bibfnamefont{P.}~\bibnamefont{{Pritchett}}},
  \bibinfo{journal}{grl} \textbf{\bibinfo{volume}{35}}, \bibinfo{pages}{L22109}
  (\bibinfo{year}{2008}).

\bibitem[{\citenamefont{Shay et~al.}(2011)\citenamefont{Shay, Drake, Eastwood,
  and Phan}}]{shay2011super}
\bibinfo{author}{\bibfnamefont{M.}~\bibnamefont{Shay}},
  \bibinfo{author}{\bibfnamefont{J.}~\bibnamefont{Drake}},
  \bibinfo{author}{\bibfnamefont{J.}~\bibnamefont{Eastwood}}, \bibnamefont{and}
  \bibinfo{author}{\bibfnamefont{T.}~\bibnamefont{Phan}},
  \bibinfo{journal}{Physical review letters} \textbf{\bibinfo{volume}{107}},
  \bibinfo{pages}{065001} (\bibinfo{year}{2011}).

\bibitem[{\citenamefont{Lapenta et~al.}(2018)\citenamefont{Lapenta, Pucci,
  Olshevsky, Servidio, Sorriso-Valvo, Newman, and
  Goldman}}]{lapenta2018nonlinear}
\bibinfo{author}{\bibfnamefont{G.}~\bibnamefont{Lapenta}},
  \bibinfo{author}{\bibfnamefont{F.}~\bibnamefont{Pucci}},
  \bibinfo{author}{\bibfnamefont{V.}~\bibnamefont{Olshevsky}},
  \bibinfo{author}{\bibfnamefont{S.}~\bibnamefont{Servidio}},
  \bibinfo{author}{\bibfnamefont{L.}~\bibnamefont{Sorriso-Valvo}},
  \bibinfo{author}{\bibfnamefont{D.~L.} \bibnamefont{Newman}},
  \bibnamefont{and} \bibinfo{author}{\bibfnamefont{M.~V.}
  \bibnamefont{Goldman}}, \bibinfo{journal}{Journal of Plasma Physics}
  \textbf{\bibinfo{volume}{84}}, \bibinfo{pages}{715840103}
  (\bibinfo{year}{2018}).

\bibitem[{\citenamefont{Guzdar et~al.}(2010)\citenamefont{Guzdar, Hassam,
  Swisdak, and Sitnov}}]{guzdar2010simple}
\bibinfo{author}{\bibfnamefont{P.}~\bibnamefont{Guzdar}},
  \bibinfo{author}{\bibfnamefont{A.}~\bibnamefont{Hassam}},
  \bibinfo{author}{\bibfnamefont{M.}~\bibnamefont{Swisdak}}, \bibnamefont{and}
  \bibinfo{author}{\bibfnamefont{M.}~\bibnamefont{Sitnov}},
  \bibinfo{journal}{Geophysical Research Letters}
  \textbf{\bibinfo{volume}{37}}, \bibinfo{pages}{L20102}
  (\bibinfo{year}{2010}).

\bibitem[{\citenamefont{Lapenta et~al.}(2016)\citenamefont{Lapenta, Goldman,
  Newman, and Markidis}}]{lapenta2016energy}
\bibinfo{author}{\bibfnamefont{G.}~\bibnamefont{Lapenta}},
  \bibinfo{author}{\bibfnamefont{M.~V.} \bibnamefont{Goldman}},
  \bibinfo{author}{\bibfnamefont{D.~L.} \bibnamefont{Newman}},
  \bibnamefont{and} \bibinfo{author}{\bibfnamefont{S.}~\bibnamefont{Markidis}},
  \bibinfo{journal}{Plasma Physics and Controlled Fusion}
  \textbf{\bibinfo{volume}{59}}, \bibinfo{pages}{014019}
  (\bibinfo{year}{2016}).

\end{thebibliography}
\end{document}